\documentclass[11pt]{article}

\usepackage{calc}
\usepackage{color}
\usepackage{graphicx}
\usepackage{float}
\usepackage{placeins}
\usepackage{url}

\usepackage{booktabs}
\usepackage{tabularx}
\usepackage{multirow}
\usepackage{makecell}
\usepackage{threeparttable}
\usepackage{wrapfig}
\usepackage[export]{adjustbox}

\usepackage{mathtools}
\usepackage{bm}
\usepackage[bbgreekl]{mathbbol}
\usepackage{siunitx}

\usepackage{algorithmic}
\usepackage{algorithm}

\usepackage[superscript,biblabel]{cite}
\usepackage{titleref}

\usepackage[left=0.5in,right=0.5in,top=0.5in,bottom=0.5in,includefoot,heightrounded]{geometry}
\usepackage{lscape}
\usepackage[margin=15pt,font=small,labelfont={bf,sf},justification=justified]{caption}
\usepackage{subcaption} 
\usepackage{titling}
\usepackage{authblk}
\usepackage[pagewise]{lineno}
\usepackage{soul}
\usepackage{blindtext}
\usepackage{lipsum}
\usepackage{easylist}
\usepackage{tikz}

\usepackage{hyperref}
\usepackage{cleveref}

\newif\ifbembo
\newif\ifcharter
\newif\iferewhon
\newif\iflibertine
\newif\iflibertinealt
\newif\ifpalantino
\newif\iftimesnewroman

\bembofalse
\chartertrue
\erewhonfalse
\libertinefalse
\libertinealtfalse
\palantinofalse
\timesnewromanfalse

\ifbembo
  \usepackage[p,osf]{ETbb} 
  \usepackage[scaled=.95,type1]{cabin} 
  \usepackage[varqu,varl]{zi4}
  \usepackage[T1]{fontenc}
  \usepackage[libertine,vvarbb]{newtxmath}
  \usepackage[cal=boondoxo,bb=boondox]{mathalfa}
\fi

\ifcharter
  \usepackage[scaled=.98,p]{XCharter}
  \usepackage[scaled=1.04,varqu,varl]{inconsolata}
  \usepackage[type1]{cabin}
  \usepackage[uprightscript,xcharter,vvarbb,scaled=1.05]{newtxmath}
  \usepackage[cal=euler,scr=boondoxo,bb=boondox]{mathalpha}  
\fi

\iferewhon
  \usepackage[p,osf,scaled=.98,space]{erewhon} 
  \usepackage[varqu,varl]{inconsolata} 
  \usepackage[type1,scaled=.95]{cabin} 
  \usepackage[utopia,vvarbb]{newtxmath}
\fi

\iflibertine
  \usepackage[sb]{libertinus}
  \usepackage[T1]{fontenc}
  \usepackage{textcomp}
  \usepackage[varqu,varl]{zi4}
  \usepackage{libertinust1math} 
  \usepackage[scr=boondoxo,bb=boondox]{mathalpha} 
\fi

\iflibertinealt
  \usepackage[sb]{libertinus}
  \usepackage[T1]{fontenc}
  \usepackage{textcomp}
  \usepackage{libertinust1math} 
  \usepackage[scr=boondoxo,bb=boondox]{mathalpha} 
\fi

\ifpalantino
  \usepackage[largesc]{newpxtext}
  \usepackage[vvarbb]{newpxmath}
\fi

\iftimesnewroman
  \usepackage{newtxtext} 
  \usepackage[scaled=0.95]{inconsolata} 
  \usepackage[vvarbb]{newtxmath} 
\fi

\usepackage{microtype}
\pretitle{\begin{center}\bfseries\Large}
\posttitle{\end{center}}

\predate{\begin{center}\normalsize}
\postdate{\end{center}}

\pretitle{\begin{center}\bfseries\sffamily\LARGE}
\posttitle{\end{center}}
\usepackage{abstract}

\allowdisplaybreaks
\usepackage{sectsty} 
\allsectionsfont{\sffamily}

\makeatletter
\patchcmd{\LS@rot}{90}{-90}{}{}
\patchcmd{\endlandscape}{90}{-90}{}{}
\makeatother

\renewcommand{\vec}[1]{\bm{\mathrm{#1}}}

\def \L2{L^2}

\definecolor{vargreen}{rgb}{0.0, 0.5, 0.0}
\definecolor{varpurp}{rgb}{0.5, 0.0, 0.5}

\title{
  Fully GPU-Accelerated Immersed Boundary Method for Fluid-Structure Interaction in Complex Cardiac Models  
}
\author[1-3]{Pengfei ~Ma} 
\author[1-3$^{*}$]{Li ~Cai}
\author[1-3]{Xuan ~Wang}
\author[4]{Xiaoyu ~Luo}
\author[4]{Hao ~Gao}

\affil[1]{Xi'an Key Laboratory of Scientific Computation and Applied Statistics, China}
\affil[2]{NPU-UoG International Cooperative Lab for Computation and Application in Cardiology, China}
\affil[3]{School of Mathematics and Statistics, Northwestern Polytechnical University, China}
\affil[4]{School of Mathematics and Statistics, University of Glasgow, UK}
\affil[*]{\texttt{Corresponding author, caili@nwpu.edu.cn}}

\date{} 

\begin{document}

\maketitle

\begin{abstract}
    Fluid-structure interaction (FSI) plays a crucial role in cardiac mechanics, 
    where the strong coupling between fluid flow and deformable structures presents 
    significant computational challenges. The immersed boundary (IB) method 
    efficiently handles large deformations and contact without requiring mesh 
    regeneration. However, solving complex FSI problems demands high computational 
    efficiency, making GPU acceleration essential to leverage massive parallelism, high throughput, and memory bandwidth.  
    We present a fully GPU-accelerated algorithm for the IB method to solve FSI problems in complex cardiac models. 
    The Navier-Stokes equations are discretized using the finite difference method, 
    while the finite element method is employed for structural mechanics. 
    Traditionally, IB methods are not GPU-friendly due to irregular memory access and limited parallelism. 
    The novelty of this work lies in eliminating sparse matrix storage and operations entirely, 
    significantly improving memory access efficiency and fully utilizing GPU computational capability. 
    Additionally, the structural materials can be modeled using general hyperelastic constitutive laws, 
    including fiber-reinforced anisotropic biological tissues such as the Holzapfel-Ogden (HO) model.  
    Furthermore, a combined geometric multigrid solver is used to accelerate the convergence. 
    The full FSI system, consisting of millions of degrees of freedom, achieves a per-timestep computation time of just 0.1 seconds.  
    We conduct FSI simulations of the left ventricle, mitral valve, and aortic valve, achieving results with high consistency. 
    Compared to single-core CPU computations, our fully GPU-accelerated approach delivers speedups exceeding 100 times.
\end{abstract}   


\newpage

\section{Introduction}
\label{sec:introduction}

The plan of this article is as follows. In Section 2, we provide an overview of 
the Navier-Stokes equations, immersed boundary formulations, and the finite element method, 
discussing each independently. In Section 3, we present the mathematical formulation for coupling these methods, 
highlighting the benefits of the three adaptations on algorithms optimized for GPU architecture. 
To validate the accuracy and efficiency of the proposed approach, we conduct four test cases in Section 4. 
These include Cook's membrane, intraventricular flow, aortic valve, and mitral valve simulations. 
We employ complex constitutive models, such as the Holzapfel–Ogden model, which incorporates fiber-reinforced materials. 
Finally, in Section 5, we summarize our findings and highlight the strengths of our approach in capturing complex physics and efficiently handling large-scale FSI simulations.

\section{Continuous equations of motion}
\label{sec:cts_eqs_of_motion}

In the framework of the IB method, the continuous formulation of the FSI system 
describes the coupled dynamics between the fluid and the immersed structure within a fixed domain. 
Let $\Omega$ be the computational domain, and $B_{t}$ be the region occupied by the structure at time $t$. 
The computational domain uses fixed Eulerian coordinates $\mathbf{x}=(x_{1},x_{2},x_{3})\in\Omega$, while the reference configuration 
of the structure employs Lagrangian reference coordinates $\mathbf{X}=(X_{1},X_{2},X_{3}) \in B_{r}$, where $B_{r}$ is the region occupied by the structure at time $t=0$. 
The transformation from the reference region $B_{r}$ to the current region $B_{t}$ is described by a bijection $\mathcal{X}(\mathbf{X}, t): B_{r} \mapsto B_{t}$, 
which characterizes the deformation of the structure, such that
\begin{align*}
    \mathbf{x}=\mathcal{X}(\mathbf{X},t)~~~~\text{for all}~\mathbf{X}\in B_{r}.
\end{align*}
Thus, the deformation gradient tensor is defined as:
\begin{align*}
    \mathbb{F}(\mathbb{X},t)=\frac{\partial \mathcal{X}}{\partial \mathbf{X}}(\mathbf{X},t),
\end{align*}
and $J(\mathbf{X},t)=\text{det}(\mathbb{F}(\mathbf{X},t))$ is the determinant of $\mathbb{F}(\mathbf{X},t)$.

Within the Eulerian formulation, the Cauchy stress tensor $\boldsymbol{\sigma}(\mathbf{x},t)$ of flexible structures in an FSI system is expressed as:
 \begin{align*}
    \boldsymbol{\sigma}(\mathbf{x}, t) = \boldsymbol{\sigma}^f(\mathbf{x}, t) + 
    \begin{cases} 
        \boldsymbol{\sigma}^e(\mathbf{x}, t), & \mathbf{x} \in B_t, \\
        0, & \mathbf{x} \in \Omega / B_t,
    \end{cases}
 \end{align*}
where $\boldsymbol{\sigma}^e(\mathbf{x}, t)$ is the elastic component of the stress tensor that exists only in the solid domain, 
and $\boldsymbol{\sigma}^f(\mathbf{x}, t)$ is the fluid-like component of the stress tensor that exists everywhere in $\Omega$.
To describe the elastic properties of the structure in the Lagrangian coordinate system, $\boldsymbol{\sigma}^e(\mathbf{x}, t)$ 
is written as 
\begin{align*}
    \boldsymbol{\sigma}^e(\mathbf{x}, t)=\boldsymbol{\sigma}^e(\mathcal{X}(\mathbf{X},t), t)=J^{-1}(\mathbf{X},t)\mathbb{P}(\mathbf{X},t)\mathbb{F}^{T}(\mathbf{X},t),
\end{align*}
where $\mathbb{P}(\mathbf{X},t)$ represents the first Piola-Kirchhoff stress tensor. For the hyperelastic constitutive model considered here, we determine $\mathbb{P}(\mathbf{X},t)$ from a strain energy function $W(\mathbb{F}(\mathbf{X},t))$ by
\begin{align*}
    \mathbb{P}(\mathbf{X},t)=\frac{\partial W(\mathbb{F}(\mathbf{X},t))}{\partial \mathbb{F}(\mathbf{X},t)}.
\end{align*}
Given that we consider the fluid to be Newtonian, then $\boldsymbol{\sigma}^f(\mathbf{x}, t)$ is 
\begin{align*}
    \boldsymbol{\sigma}^f(\mathbf{x}, t)=-p(\mathbf{x},t)\mathbb{I}+\mu\left[\nabla\mathbf{u}(\mathbf{x},t)+(\nabla\mathbf{u}(\mathbf{x},t))^{T}\right].
\end{align*}
Here, $\mathbf{u}(\mathbf{x}, t)$ represents the Eulerian velocity field, and $p(\mathbf{x}, t)$ represents the pressure field, 
which also acts as the Eulerian Lagrange multiplier to enforce the incompressibility constraint $\nabla \cdot \mathbf{u} = 0$,
$\mathbb{I}$ denotes the identity tensor, and $\mu$ is the dynamic viscosity.

Eulerian and Lagrangian variables are coupled through integral transforms with delta function kernels in the IB framework. 
Let $\delta(\mathbf{x})=\Pi^{d}_{i=1}\delta(x_{i})$ denote the $d$-dimensional delta function, where $x_{i}$
are the Cartesian components of the vector $\mathbf{x}$ \cite{peskin2002immersed, strichartz2003guide}. Let $\mathbf{f}(\mathbf{x},t)$ be the Eulerian elastic force density, 
$\mathbf{F}(\mathbf{X},t)$ be the Lagrangian elastic force density, and $\mathbf{U}(\mathbf{X},t)$ be the Lagrangian velocity field.
Based on the definitions of the interpolation operator $\mathcal{I}(\cdot,\cdot)$ and spreading operator $\mathcal{E}(\cdot,\cdot)$ provided in \cite{ma2024unconditionally}, 
we establish the relationships between $\mathbf{F}(\mathbf{X},t)$ and $\mathbf{f}(\mathbf{x},t)$, 
as well as $\mathbf{u}(\mathbf{x},t)$ and $\mathbf{U}(\mathbf{X},t)$ as follows:
\begin{align}
    \mathbf{f}(\mathbf{x}, t)=\mathcal{E}(\mathcal{X},\mathbf{F})=\int_{B_{r}}\mathbf{F}(\mathbf{X},t)\delta(\mathbf{x}-\mathcal{X}(\mathbf{X},t))\ d\mathbf{X},\\
    \mathbf{U}(\mathbf{X},t)=\mathcal{I}(\mathcal{X},\mathbf{u})=\int_{\Omega}\mathbf{u}(\mathbf{x},t)\delta\left(\mathbf{x}-\mathcal{X}(\mathbf{X},t)\right) d\mathbf{x}.
\end{align}

The present study is based on the IB method derived by Boffi et al.\,\cite{boffi2008hyper},
find $\mathbf{u}(\mathbf{x},t)$, $p(\mathbf{x},t)$, $\mathcal{X}(\mathbf{x},t)$ which satisfy
\begin{align} \label{IB::continuous::equations}
        &\rho \frac{D \mathbf{u}}{D t}(\mathbf{x}, t)+\nabla p(\mathbf{x}, t)-\mu \Delta \mathbf{u}(\mathbf{x}, t)=\mathbf{f}(\mathbf{x}, t),  & \text { in } \Omega \times[0, T], \nonumber\\
        &\nabla \cdot \mathbf{u}(\mathbf{x}, t)=0, & \text { in } \Omega \times[0, T], \nonumber \\
        &\int_{B_r} \mathbf{F}(\mathbf{X}, t) \cdot \mathbf{V}(\mathbf{X}) d\mathbf{X}=-\int_{B_r} \mathbb{P}(\mathbf{X}, t): \nabla_{\mathbf{X}} \mathbf{V}(\mathbf{X}) d\mathbf{X}, & \text { in } B_{r} \times[0, T], \nonumber\\
        &\mathbf{f}(\mathbf{x}, t)=\mathcal{E}(\mathcal{X}(\mathbf{X},t),\mathbf{F}(\mathbf{X},t)), & \text { in } B_{r} \times[0, T], \nonumber\\
        &\frac{\partial \mathcal{X}(\mathbf{X}, t)}{\partial t}=\mathbf{U}(\mathbf{X},t)=\mathcal{I}(\mathcal{X}(\mathbf{X},t),\mathbf{u}(\mathbf{x},t)), & \text { in } \Omega \times[0, T],  
\end{align}
where $\rho$ be the density of the FSI system. And imposed the initial and boundary conditions
\begin{equation}\label{eq::IB::continuous::boundary::intial}
    \begin{cases}\mathbf{u}(\mathbf{x}, t)=\mathbf{w}(\mathbf{x},t), & \text { in } \partial \Omega \times[0, T], \\ \mathbf{u}(\mathbf{x}, 0)=\mathbf{u}_0(\mathbf{x}), & \text { in } \Omega .\end{cases}
\end{equation}
Here, $\frac{D\mathbf{u}}{Dt}(\mathbf{x},t)=\frac{\partial \mathbf{u}}{\partial t}(\mathbf{x},t)+(\mathbf{u}(\mathbf{x},t)\cdot\nabla)\mathbf{u}(\mathbf{x},t),$ $\mathbf{V}(\mathbf{X})$ is any smooth function.

\section{Constitutive laws}
The Cauchy stress of the immersed elastic structure is decomposed in the traditional IB method as \cite{boffi2008hyper, griffith2017hybrid, li2024local}:
 \begin{align}\label{constitutive_laws::stress}
    \boldsymbol{\sigma} =\boldsymbol{\sigma}^{v}-p\mathbb{I}+
    \begin{cases}
    0, & \mathbf{x} \in \Omega_t^{f}, \\
    \boldsymbol{\sigma}^{e}, & \mathbf{x} \in \Omega_t^s.  
    \end{cases}
 \end{align}
where $\boldsymbol{\sigma}^{v}=\mu(\nabla\mathbf{v}+\nabla\mathbf{v}^{T})$ is the deviatoric viscous stress and $\boldsymbol{\sigma}^{e}$ is the elastic stress.
To address the poor numerical results \cite{vadala2020stabilization} that may arise from Eq.\,\eqref{constitutive_laws::stress}, $\boldsymbol{\sigma}^{e}$ is decomposed as:
$$\boldsymbol{\sigma}^{e}=\text{dev}[\boldsymbol{\sigma}^{e}]-\pi_{\text{stab}}\mathbb{I},$$
where $\text{dev}[\boldsymbol{\sigma}^{e}]$ represents the deviatoric stress component, and $\pi_{\text{stab}}\mathbb{I}$ represents the volumetric stress component, which serves as a stabilization term that acts like an additional pressure in solid region.
Based on \cite{vadala2020stabilization}, $\pi_{\text{stab}}$ depends on the volumetric energy $U(J)$, which is solely related to the volume changes of the structure.
\begin{align*}
    \pi_{\text{stab}}=-\frac{\partial U(J)}{\partial J}.
\end{align*}
In this paper, a simple form of $U(J)$ is chosen as $U(J)=\frac{\kappa_{\text{stab}}}{2}(\text{ln}J)^{2}$, where the value of $\kappa_{\text{stab}}$ depends on the numerical Poisson ratio $\nu_{\text{stab}}$,
and $\kappa_{\text{stab}}=\frac{2G_{T}(1+\nu_{\text{stab}})}{3(1-2\nu_{\text{stab}})}$.
Thus, for hyperelastic materials, based on $\boldsymbol{\sigma}^{e}=\frac{1}{J}\frac{\partial\Psi(\mathbb{F})}{\partial \mathbb{F}}\mathbb{F}^{T}$, it follows that:
\begin{align*}
    \Psi(\mathbb{F}) = W(\mathbb{F})+U(J),
\end{align*}
where $W(\mathbb{F})$ is the volume-preserving components, $U(J)$ is the volume-changing component.

To ensure material frame invariance, this study considers anisotropic materials and, based on \cite{vadala2020stabilization,bonet1997nonlinear}, 
defines their strain energy function $\Psi$ in the following general form using the right Cauchy-Green tensor $\mathbb{C}=\mathbb{F}^{T}\mathbb{F}$:  
\begin{align*}
 \Psi = W(I_{1},I_{4f},I_{4s},I_{5f}, I_{8fs})+U(J),
 \end{align*}
 where $I_{1}=\text{tr}(\mathbb{C})$, $I_{4f} = \mathbf{e}_{f}\cdot(\mathbb{C}\mathbf{e}_{f})$, $I_{5f} = \mathbf{e}_{f}\cdot(\mathbb{C}^{2}\mathbf{e}_{f})$, $I_{4s} = \mathbf{e}_{s}\cdot(\mathbb{C}\mathbf{e}_{s})$, $I_{8fs} = \mathbf{e}_{f}\cdot(\mathbb{C}\mathbf{e}_{s})$, $\mathbf{e}_{f}$ represents the fiber direction in the reference configuration, and $\mathbf{e}_{s}$ represents the sheet direction in the reference configuration.
 
Next, to remove volume change information, the material used in this study introduces modified invariants based on $I_1$, as follows:
 \begin{align*}
     \bar{I}_{1}=J^{-2/3}I_{1}.
 \end{align*}
Therefore, the $\Psi$ represented by the modified invariant $\bar{I}_1$ is:
\begin{align*}
    \Psi = \bar{W}(\bar{I}_{1},I_{4f},I_{4s},I_{5f}, I_{8fs})+U(J).
\end{align*} 

Finally, we will briefly introduce several common constitutive models, including their free energy functions and the corresponding first Piola-Kirchhoff stress tensors.
\begin{itemize}
    \item Modified standard reinforcing model \cite{vadala2020stabilization}:
    \begin{align}
        \Psi_1&=\frac{G_{T}}{2}(\bar{I}_{1}-3)+\frac{G_{T}-G_{L}}{2}(2I_{4f}-I_{5f}-1) \nonumber\\
        &+\frac{E_{L}+G_{T}-4G_{L}}{8}(I_{4f}-1)^{2}+\frac{\kappa_{\text{stab}}}{2}(\text{ln}J)^{2},\\
        \mathbb{P}_{1}&=G_{T}J^{-2/3}(\mathbb{F}-\frac{I_{1}}{3}\mathbb{F}^{-T})+(G_{T}-G_{L})(2\mathbb{F}\mathbf{e}_{f}\otimes \mathbf{e}_{f}-\mathbb{F}\mathbf{e}_{f}\otimes \mathbf{e}_{f}\mathbb{C}-\mathbb{F}\mathbb{C}\mathbf{e}_{f}\otimes \mathbf{e}_{f}) \nonumber\\
        &+\frac{E_{L}+G_{T}-4G_{L}}{2}(I_{4f}-1)\mathbb{F}\mathbf{e}_{f}\otimes \mathbf{e}_{f}+\kappa_{\text{stab}}\text{ln}(J)\mathbb{F}^{-T},  
    \end{align}
    where $G_{T}$, $G_{L}$, $E_{L}$ are material constants.
    \item Fiber-reinforced hyperelastic model 1\cite{cai2019some}:
    \begin{align}
        \Psi_2&=C_{1}(\bar{I}_{1}-3)+\frac{a_{f}}{2b_{f}}\left(\text{exp}\left[b_{f}(I_{4f}^{*}-1)^{2}\right]-1\right)+\frac{\kappa_{\text{stab}}}{2}(\text{ln}J)^{2},\\
        \mathbb{P}_{2}&=2C_{1}J^{-2/3}(\mathbb{F}-\frac{I_1}{3}\mathbb{F}^{-T})+2a_f(I_{4f}^{*}-1)\text{exp}\left[b_f(I^{*}_{4f}-1)^{2}\right]\mathbb{F}\mathbf{e}_{f}\otimes \mathbf{e}_{f}+\kappa_{\text{stab}}\text{ln}(J)\mathbb{F}^{-T},
    \end{align}
    where $I^{*}_{4f}=\text{max}(I_{4f},1)$, $C_{1}$, $a_{f}$, $b_{f}$ are the material constants.
    \item Fiber-reinforced hyperelastic model 2\cite{cai2021comparison}:
    \begin{align}
        \Psi_3 &= C_{10}\left(\text{exp}\left[C_{01}(\bar{I}_{1}-3)\right]-1\right)+\frac{a_{f}}{2b_{f}}\left(\text{exp}\left[b_{f}(I_{4f}-1)^{2}\right]-1\right)+\frac{\kappa_{\text{stab}}}{2}(\text{ln}J)^{2},\\
        \mathbb{P}_{3}&=2C_{10}C_{01}\text{exp}\left[C_{10}(\bar{I}_{1}-3)\right]J^{-2/3}\left(\mathbb{F}-\frac{I_{1}}{3}\mathbb{F}^{-T}\right)+2a_{f}(I_{4f}-1)\text{exp}\left[b_{f}(I_{4f}-1)^{2}\right]\mathbb{F}\mathbf{e}_{f}\otimes \mathbf{e}_{f}+\kappa_{\text{stab}}\text{ln}(J)\mathbb{F}^{-T},
        \end{align}
    where $C_{10}$, $C_{01}$ $a_{f}$, $b_{f}$ are the material constants.
    \item Holzapfel-Ogden (HO) model \cite{thekkethil2023stabilized}:
    \begin{align}
        \Psi_{4}&=\frac{a}{2b}\text{exp}[b(\bar{I}_{1}-3)]+\sum_{i=f,s}\frac{a_{i}}{2b_{i}}\left\{ \text{exp}\left[b_{i}(I_{4i}^{*}-1)^{2}\right]-1\right\}\\
        &+\frac{a_{fs}}{2b_{fs}}\{\text{exp}\left[b_{fs}(I_{8fs})^{2}\right]-1\}+\frac{\kappa_{\text{stab}}}{2}(\text{ln}J)^{2},\\
        \mathbb{P}_{4}&=a\text{exp}\left[b(\bar{I}_{1}-3)\right]J^{-2/3}(\mathbb{F}-\frac{I_1}{3}\mathbb{F}^{T})+\sum_{i=f,s}2a_{i}(I_{4i}^{*}-1)\text{exp}\left[b_{i}(I_{4i}^{*}-1)^{2}\right]\mathbb{F}\mathbf{e}_{i}\otimes\mathbf{e}_{i}\\
        &+2a_{fs}I_{8fs}\text{exp}\left[b_{fs}(I_{8fs})^{2}\right]\mathbb{F}\mathbf{e}_{f}\otimes \mathbf{e}_{s}+\kappa_{\text{stab}}\text{ln}(J)\mathbb{F}^{-T},
    \end{align}
    where $I_{4i}^{*}=\text{max}(I_{4i},1)$, $a$, $b$, $a_{i}$, $b_{i}$, $a_{fs}$, and $b_{fs}$ are material constants, with $i=f,s$.
\end{itemize}

\section{Numerical Discretization}

\subsection{Temporal Discretization}

For system \eqref{IB::continuous::equations}, we begin by considering the temporal semi-discretization. The time interval 
$[0,T]$ is partitioned into $N$ uniformly spaced, non-overlapping subintervals $(t_{n}, t_{n+1}]$, where $n=1,2,...,N$, 
with a constant time step size $t_{n}-t_{n-1}=\Delta t$. Here, $(\cdot)^{n}$ denotes the value of $(\cdot)$ at time $t_{n}$.
In this work, the backward Euler time-stepping scheme is employed for the temporal discretization of system \eqref{IB::continuous::equations} from \cite{ma2024unconditionally}, 
and the resulting semi-discrete system is given as follows:
\begin{align}
    \rho(\frac{\mathbf{u}^{n+1}-\mathbf{u}^{n}}{\Delta t}+\mathbf{u}^{n}\cdot\nabla\mathbf{u}^{n+1})-\mu\Delta\mathbf{u}^{n+1}+\nabla p^{n+1}&=\mathbf{f}^{n+1},\quad~&\text{in}~\Omega, \nonumber\\
    \nabla\cdot\mathbf{u}^{n+1}&=0,\quad~&\text{in}~\Omega, \nonumber\\
    \int_{B_{r}}\mathbf{F}^{n+1}\cdot \mathbf{V}(\mathbf{X})d\mathbf{X}&=-\int_{B_{r}}\mathbb{P}^{n+1}:\nabla_{\mathbf{X}}\mathbf{V}(\mathbf{X})d\mathbf{X},\quad~&\text{in}~B_{r} \nonumber\\
    \mathbf{f}^{n+1}&=\mathcal{E}(\mathcal{X}^{n},\mathbf{F}^{n+1}),\quad~&\text{in}~B_{r}, \nonumber\\
    \frac{\mathcal{X}^{n+1}-\mathcal{X}^{n}}{\Delta t}&=\mathcal{I}(\mathcal{X}^{n},\mathbf{u}^{n}),\quad~&\text{in}~\Omega.
\end{align}

System \eqref{IB::continuous::equations} employs an explicit coupling method to handle fluid-structure interactions. Although a large number of implicit time-stepping methods 
have been proposed, efficient, fast, and general-purpose implicit solvers remain scarce. As a result, explicit coupling methods are widely adopted in large-scale numerical simulations 
due to their simplicity and computational efficiency. Building on this advantage, we will further develop a highly efficient IB method based on a fully GPU-accelerated algorithm.

\subsection{Spatial Discretization}

In the context of spatial discretization for this explicit numerical scheme, we employ the approach developed by Griffith and Luo \cite{griffith2017hybrid},

which is demonstrated in Ref.\cite{ma2024unconditionally}.
Here, the Navier-Stokes equations are decomposed into a Helmholtz equation and a Poisson equation using the Chorin's projection method.

Given $\mathbf{u}^{n}\in \mathbf{W}_{h}$, $\mathcal{X}^{n}\in \mathbf{V}_{h}$, find $\mathbf{u}^{n+1}\in\mathbf{W}_{h}$, $p^{n+1}\in Q_{h}$, $\mathcal{X}^{n+1}\in\mathbf{V}_{h}$ which satisfy
\begin{align}
    \frac{\rho}{\Delta t}(\mathbf{u}^{*}-\mathbf{u}^{n})-\mu\Delta_{h}\mathbf{u}^{*}=\mathbf{f}^{n+1},\\
    \Delta_{h}p^{n+1}=\frac{1}{\Delta t}\nabla_{h}\cdot\mathbf{u}^{*},\\
\mathbf{u}^{n+1}=\mathbf{u}^{*}-\Delta t\nabla_{h} p^{n+1},\\
\int_{B_{r}}\mathbf{F}_{h}^{n+1}\cdot\mathbf{V}(\mathbf{X})\ d\mathbf{X}=-\int_{B_{r}}\mathbb{P}_{h}^{n+1}:\nabla_{\mathbf{X}}\mathbf{V}(\mathbf{X})\ d\mathbf{X},\\
\mathbf{f}^{n+1}=\int_{B_{r}}\mathbf{F}_{h}^{n+1}\delta(\mathbf{x}-\mathcal{X}(\mathcal{X}_{h}^{n}))\ d\mathbf{X},\\
\frac{\mathcal{X}_{h}^{n+1}-\mathcal{X}_{h}^{n}}{\Delta t}=\int_{\Omega}\mathbf{u}^{n+1}\delta(\mathbf{x}-\mathcal{X}_{h}^{n})\ d\mathbf{x}
\end{align}
Here, $\mathbf{V}(\mathbf{X})$ is any function belongs to $\mathbf{V}_{h}$. $\mathbf{u}^{*}$ is a tentative velocity. When imposed the boundary condition
\begin{equation}\label{eq::IB::discrete::boundary::intial}
    \begin{cases}\mathbf{u}^{*}=0, & \text { in } \partial \Omega, \\ \mathbf{u}^{n+1}\cdot\mathbf{n}=0, & \text { in } \partial \Omega. \end{cases}
\end{equation}
\subsection{Discretization of the Navier-Stokes equations}
For the remainder of the statement of the discretization, the Eulerian variables are discreted with the marker-and-cell staggered-grid scheme, using uniform cell of length $\Delta x$ in each coordinate direction. There are 
several prominent advantages of this staggered scheme, such as its mass conservation properties, efficiency of linear algebra, and inf-sup stability. 
\subsection{Eulerian-Lagrangian coupling operators}
Forces are transferred from the structural mesh to the Cartesian grid by  spreading. Following the discretization described in subsection 2.3 and a projection 
onto the finite element space, the force are 
\begin{align}
    f^{1}_{i+\frac{1}{2},j,k}=\int_{B_{r}}F^{1}(\mathbf{X})\delta_{h}(\mathbf{x}_{i+\frac{1}{2},j,k}-\mathcal{X}_{h}(\mathbf{X},t))d\mathbf{X}, \\
    f^{2}_{i,j+\frac{1}{2},k}=\int_{B_{r}}F^{2}(\mathbf{X})\delta_{h}(\mathbf{x}_{i,j+\frac{1}{2},k}-\mathcal{X}_{h}(\mathbf{X},t))d\mathbf{X},\\
    f^{3}_{i,j,k+\frac{1}{2}}=\int_{B_{r}}F^{3}(\mathbf{X})\delta_{h}(\mathbf{x}_{i,j,k+\frac{1}{2}}-\mathcal{X}_{h}(\mathbf{X},t))d\mathbf{X},
\end{align}
velocities are transferred from the Cartesian grid to the structural mesh by interpolation to an intermediate Lagrangian
velocity $\mathbf{U}^{IB}$:
\begin{align}
    \mathbf{U}^{IB,1}(\mathbf{X},t)=\sum_{i,j,k}u^{1}_{i+\frac{1}{2},j,k}\delta_{h}(\mathbf{x}_{i+\frac{1}{2},j,k}-\mathcal{X}_{h}(\mathbf{X},t))\Delta x^{3},\\
    \mathbf{U}^{IB,2}(\mathbf{X},t)=\sum_{i,j,k}u^{2}_{i,j+\frac{1}{2},k}\delta_{h}(\mathbf{x}_{i,j+\frac{1}{2},k}-\mathcal{X}_{h}(\mathbf{X},t))\Delta x^{3},\\
    \mathbf{U}^{IB,3}(\mathbf{X},t)=\sum_{i,j,k}u^{3}_{i,j,k+\frac{1}{2}}\delta_{h}(\mathbf{x}_{i,j,k+\frac{1}{2}}-\mathcal{X}_{h}(\mathbf{X},t))\Delta x^{3},\\
\end{align}
Note that $\mathbf{U}^{IB}$ is defined for all points in the structural mesh (in particular, at the quadrature points) but is typically not in $V^{h}$. This is the semidiscretization of Eq.\,\eqref{}. The remaining integrate equation arises from requiring that
\begin{align*}
    \int_{B_{r}}\mathbf{U}_{h}(\mathbf{X},t)\cdot\phi(\mathbf{X})d\mathbf{X}=\int_{B_{r}}\mathbf{U}^{IB}(\mathbf{X},t)\cdot\phi(\mathbf{X})d\mathbf{X},
\end{align*}
for all test functions $\phi(\mathbf{X})\in V^{h}$, i.e., $U_{h}$ is the projection of $\mathbf{U}^{IB}$ onto the finite element space. Further, note that if we discrete the integrals in Eq.\eqref{} and Eq.\,\eqref{}
with the same quadrature formula then the spreading and interpolation operators are discretely adjoint. For a through discussion on the adjointness of these two operators, see Griffith and Luo \cite{griffith2017hybrid}.

Define the nodal quadrature rule as
$\mathbb{N}_{q}=\{\mathbf{X}_{q},w_{q}\}^{m}_{q=1}$ in which $\mathbf{X}_q$ is a node of $\Gamma^{h}$ and $w_{q}=\int\phi_{q}(\mathbf{X})\cdot 1d\mathbf{X}$.

Compute the components of $\vec{\mathbf{U}}^{IB}$ by evaluating $\delta_{h}$ at each node's displaced location $\mathcal{X}_{h}(\mathbf{X}_{q})$ for each $\mathbf{X}_{q}, w_{q}\in\mathbb{N}_{q}$:
\begin{align*}
    U^{IB,1}_{q}=\sum_{i,j,k}u^{1}_{i+\frac{1}{2},j,k}\delta_{h}(\mathbf{x}_{i+\frac{1}{2},j,k}-\mathcal{X}_{h}(\mathbf{X}_{q}))\Delta x^{3}\\
    U^{IB,2}_{q}=\sum_{i,j,k}u^{2}_{i,j+\frac{1}{2},k}\delta_{h}(\mathbf{x}_{i,j+\frac{1}{2},k}-\mathcal{X}_{h}(\mathbf{X}_{q}))\Delta x^{3}\\
    U^{IB,3}_{q}=\sum_{i,j,k}u^{3}_{i,j,k+\frac{1}{2}}\delta_{h}(\mathbf{x}_{i,j,k+\frac{1}{2}}-\mathcal{X}_{h}(\mathbf{X}_{q}))\Delta x^{3}
\end{align*}

For each FE basis function compute a mean force contribution
\begin{align*}
    \vec{F}_{i}=-\sum_{(\mathbf{X}_{q},w_{q})\in\mathbb{H}_{q}}\mathbb{P}^{e}(\mathbf{X}_{q}):\nabla_{\mathbf{X}}\phi_{i}(\mathbf{X}_{q})w_{q}
\end{align*}

Spread each component of the mean force contribution at each node's displaced location $\mathcal{X}_{h}(\mathbf{X}_{q})$ as 
\begin{align*}
    f^{1}_{i+\frac{1}{2},j,k}=\sum_{(mathbf{X}_{q},w_{q}\in\mathbb{N}_{q})}\vec{F}^{1}_{q}\delta_{h}(\mathbf{x}_{i+\frac{1}{2},j,k}-\mathcal{X}_{h}(\mathbf{X}_{q}))\\
    f^{2}_{i,j+\frac{1}{2},k}=\sum_{(mathbf{X}_{q},w_{q}\in\mathbb{N}_{q})}\vec{F}^{2}_{q}\delta_{h}(\mathbf{x}_{i,j+\frac{1}{2},k}-\mathcal{X}_{h}(\mathbf{X}_{q}))\\
    f^{2}_{i,j,k+\frac{1}{2}}=\sum_{(mathbf{X}_{q},w_{q}\in\mathbb{N}_{q})}\vec{F}^{3}_{q}\delta_{h}(\mathbf{x}_{i,j,k+\frac{1}{2}}-\mathcal{X}_{h}(\mathbf{X}_{q}))
\end{align*}
The calculation yields $\mathbf{f}_{h}$.

\section{Benchmarks}

\begin{figure}[!ht]
\centering
\includegraphics[width=0.46\textwidth]{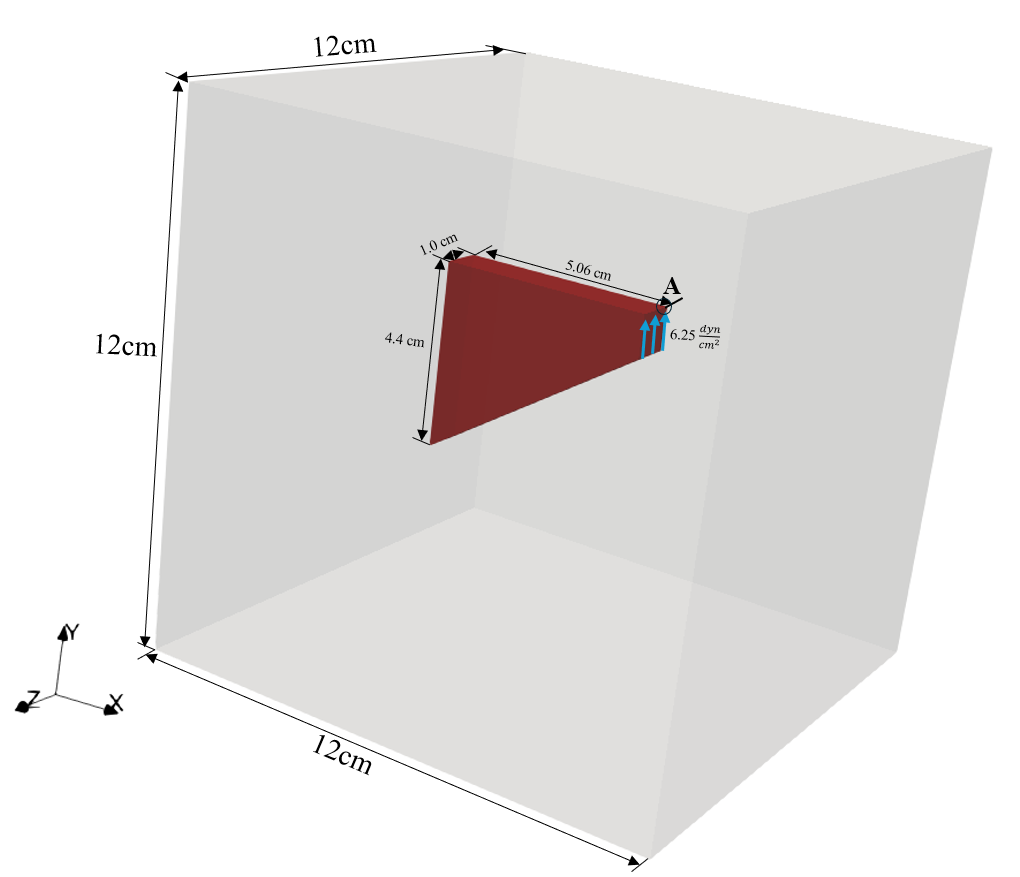}
\caption{Setup of the anisotropic Cook's membrane benchmark: The gray region represents the fixed computational domain $\Omega$, while the red region corresponds to the deformable structure.}

\label{fig::cook_membrane::anisotropic::3}
\end{figure}

The Cook’s membrane problem is a classical benchmark in incompressible elastic solid mechanics. 
To assess the computational performance of the fully GPU-implemented IB method, 
we follow the setup in \cite{vadala2020stabilization} to embed the three-dimensional anisotropic Cook's membrane into an incompressible Newtonian fluid, 
formulating a fluid-structure interaction system, 
which is then numerically solved using the IB method. The configuration of the computational domain $\Omega$, structural dimensions, and loading conditions is shown in 
Fig.\,\ref{fig::cook_membrane::anisotropic::3}.

The boundary conditions for this problem consist of a fixed left face, where constraints are enforced using a penalty parameter of $\kappa = 10^6$. 
A vertical traction of $6.25~\frac{\text{dyn}}{\text{cm}^{2}}$ is applied to the right face, increasing gradually according to the polynomial function $q(t)=-2(\frac{t}{T1})^{3}+3(\frac{t}{T1})^{2}$ and reaching its maximum at time $T_1$. 
The simulation continues until time $T_f$ to ensure the system reaches equilibrium. All other faces are subject to zero traction conditions. 
A summary of the relevant physical and numerical parameters is provided in Table \ref{tab::cook::membrane::params}.
\begin{table}[!h]
    \centering
    \begin{tabular}{lcc}
        \toprule
        \textbf{Symbol} & \textbf{Value} & \textbf{Unit}  \\
        \midrule
        $\rho$  & 1.0   & $\frac{\text{g}}{\text{cm}^3}$ \\
        \(T_1\) & 14    & $\text{s}$ \\
        \(T_f\) & 35    & $\text{s}$  \\
        $\mu$   & 0.16  & $\frac{\text{dyn}\cdot \text{s}}{\text{cm}^{2}}$\\
        \midrule
        $G_T$  & 8.0  & $\frac{\text{dyn}}{\text{cm}^2}$\\
        $G_L$  & 160.0 & $\frac{\text{dyn}}{\text{cm}^2}$ \\
        $E_L$  & 1200.0 & $\frac{\text{dyn}}{\text{cm}^2}$ \\
        $\kappa_\text{stab}$ &  112 & $\frac{\text{dyn}}{\text{cm}^2}$ \\
        $\mathbf{e}_{f}$ & $\frac{1}{\sqrt{3}}(1,1,1)$ &-\\
        \bottomrule
    \end{tabular}
    \caption{Parameters for the anisotropic Cook's membrane benchmark}
    \label{tab::cook::membrane::params}
\end{table}

In this case, the computational domain $\Omega$ is discretized using a Cartesian grid 
comprising $N = 64$ cells along each coordinate axis. A tetrahedral mesh is utilized for 
the discretization of the structure, and the forces and displacements are discretized using 
$P2$ finite elements. The degrees of freedom (DoFs) for the solid range from $m=5004$ to $m = 282,489$. 
Furthermore, the time step for the simulation is set to $\Delta t = 0.001 \text{s} $.

\begin{figure}[!ht]
    \centering
    \includegraphics[width=0.8\textwidth]{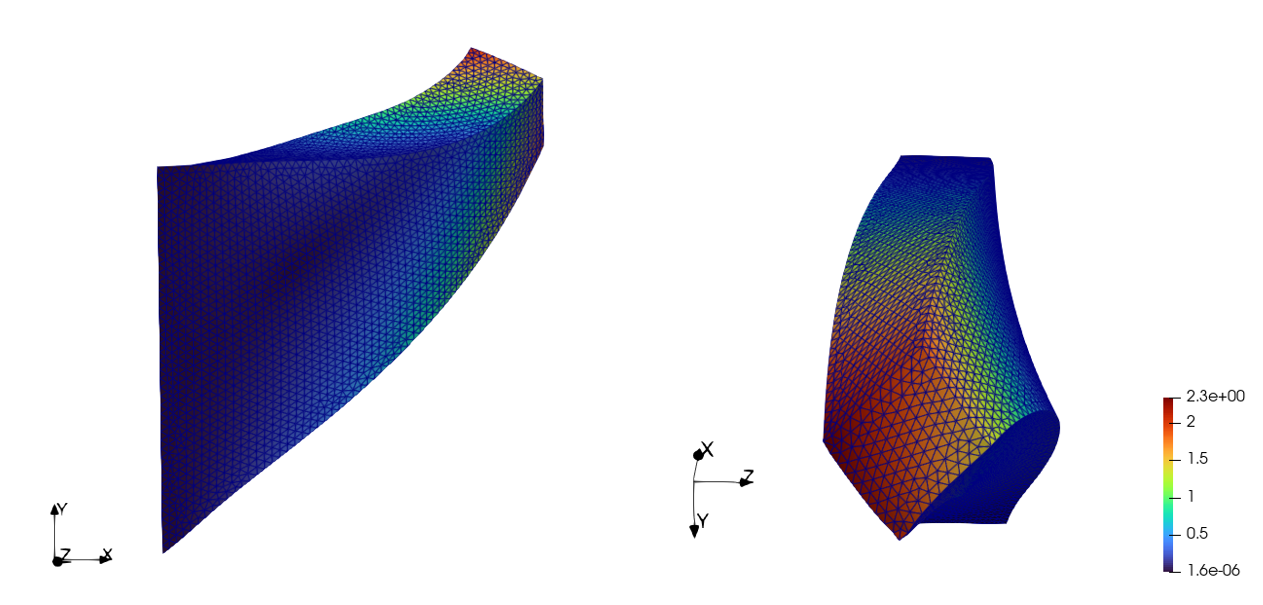}
    \caption{Deformations of the anisotropic Cook's membrane at different angles.}
    
    \label{fig::cook_membrane::anisotropic::deformation}
\end{figure}

\begin{figure}[!ht]
    \centering
    \begin{minipage}{0.45\textwidth}
        \centering
        \includegraphics[width=1.0\textwidth]{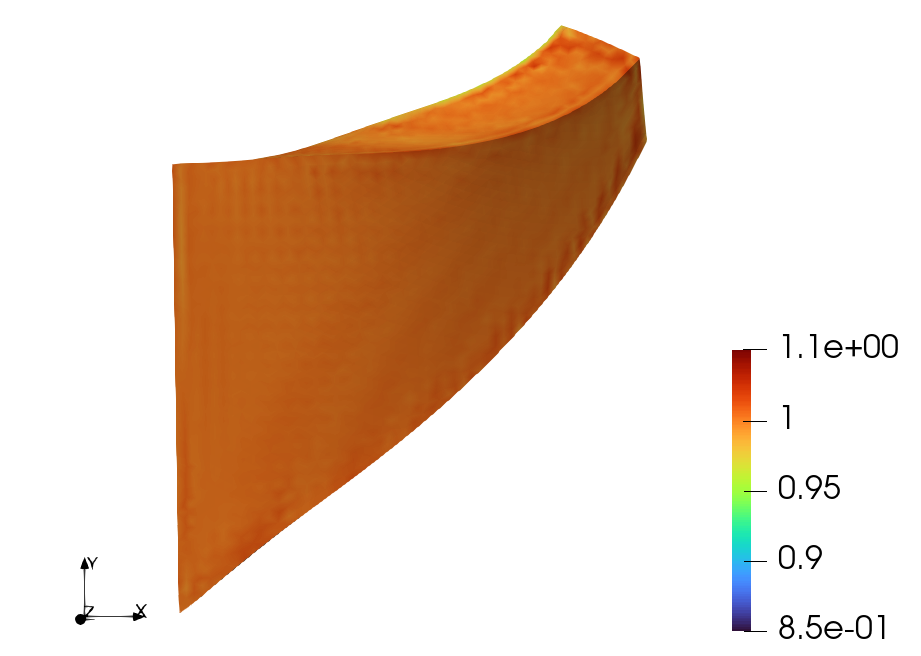}
        \subcaption{Jacobian J}
        \label{fig::cook_membrane::anisotropic::Jacobian}
    \end{minipage}
    \begin{minipage}{0.45\textwidth}
        \centering
        \includegraphics[width=1.0\textwidth]{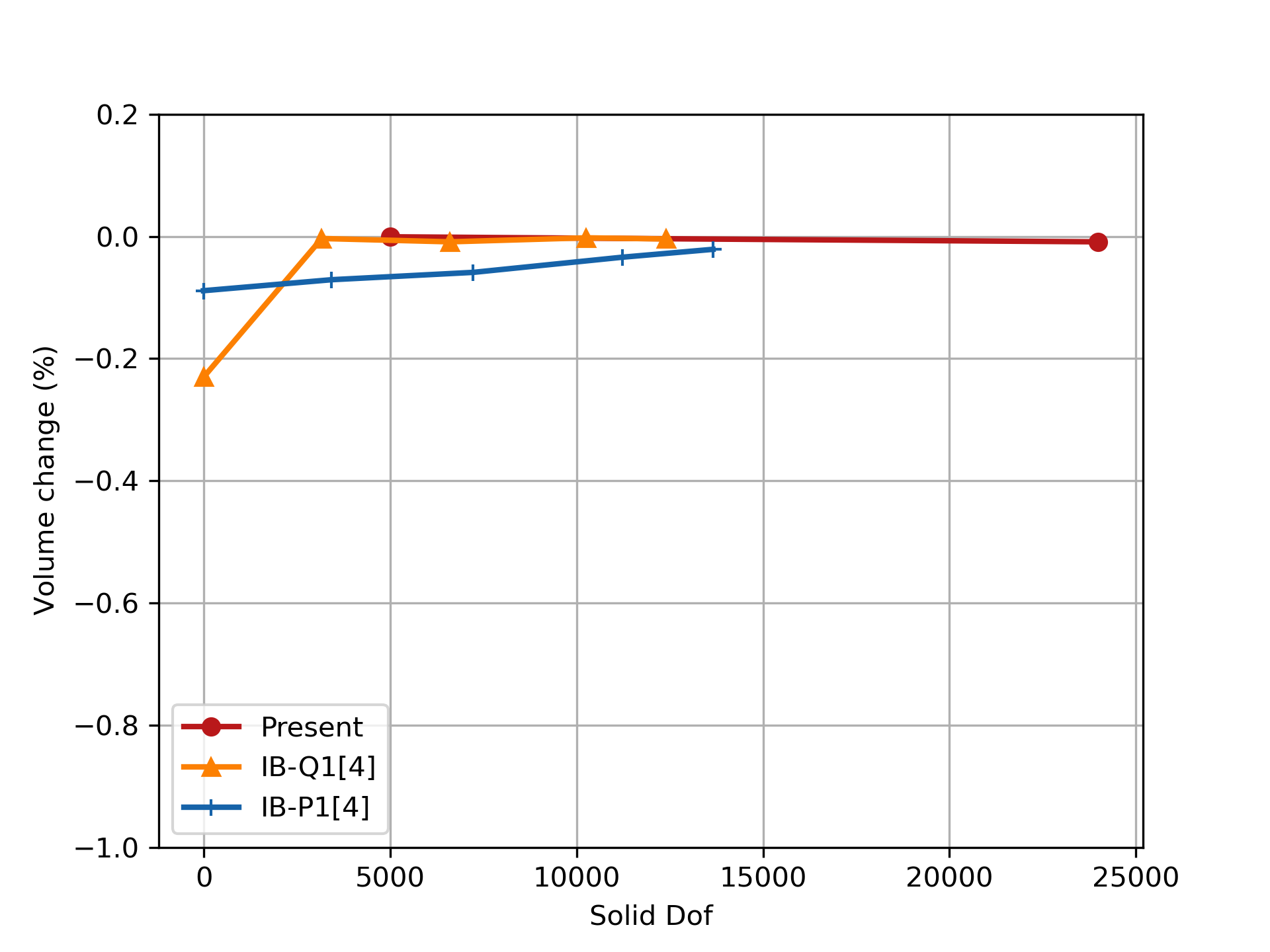}
        \subcaption{Volume change of the Cook's membrane model under Different number of Solid's Dof.}
        \label{fig::cook_membrane::anisotropic::volume_change}
    \end{minipage}
    \caption{Volume conservation for the anisotropic Cook's membrane benchmark.}
\end{figure}

\begin{figure}[!ht]
    \centering
    \begin{minipage}{0.45\textwidth}
        \centering
        \includegraphics[width=1.0\textwidth]{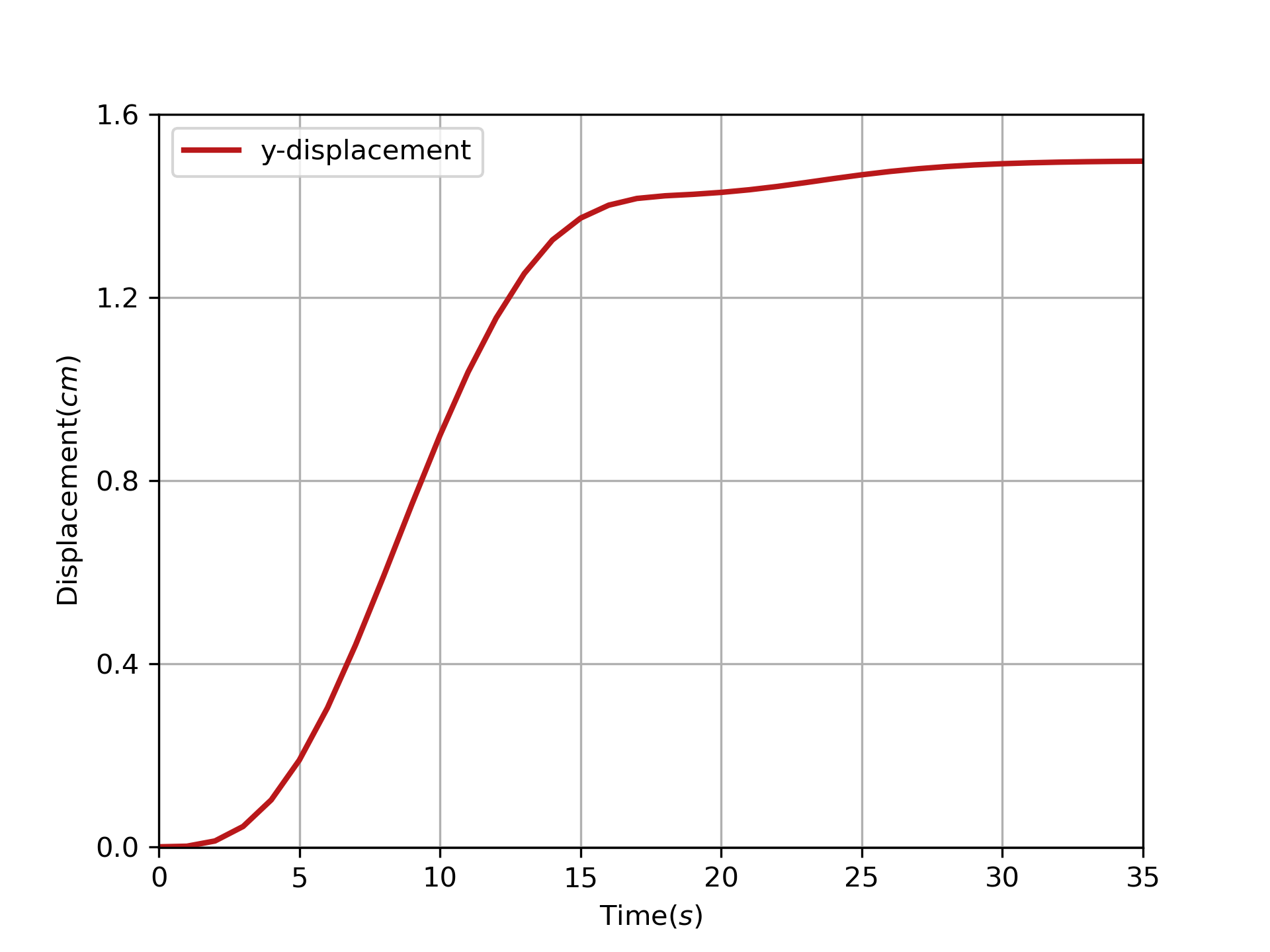}
        \subcaption{Displacement of point A over time in a solid mesh with 282,489 Nodes and a fluid mesh divided into a 64×64×64 grid.}
        \label{fig::Cook's membrane::displacement}
    \end{minipage}
    \begin{minipage}{0.45\textwidth}
        \centering
        \includegraphics[width=1.0\textwidth]{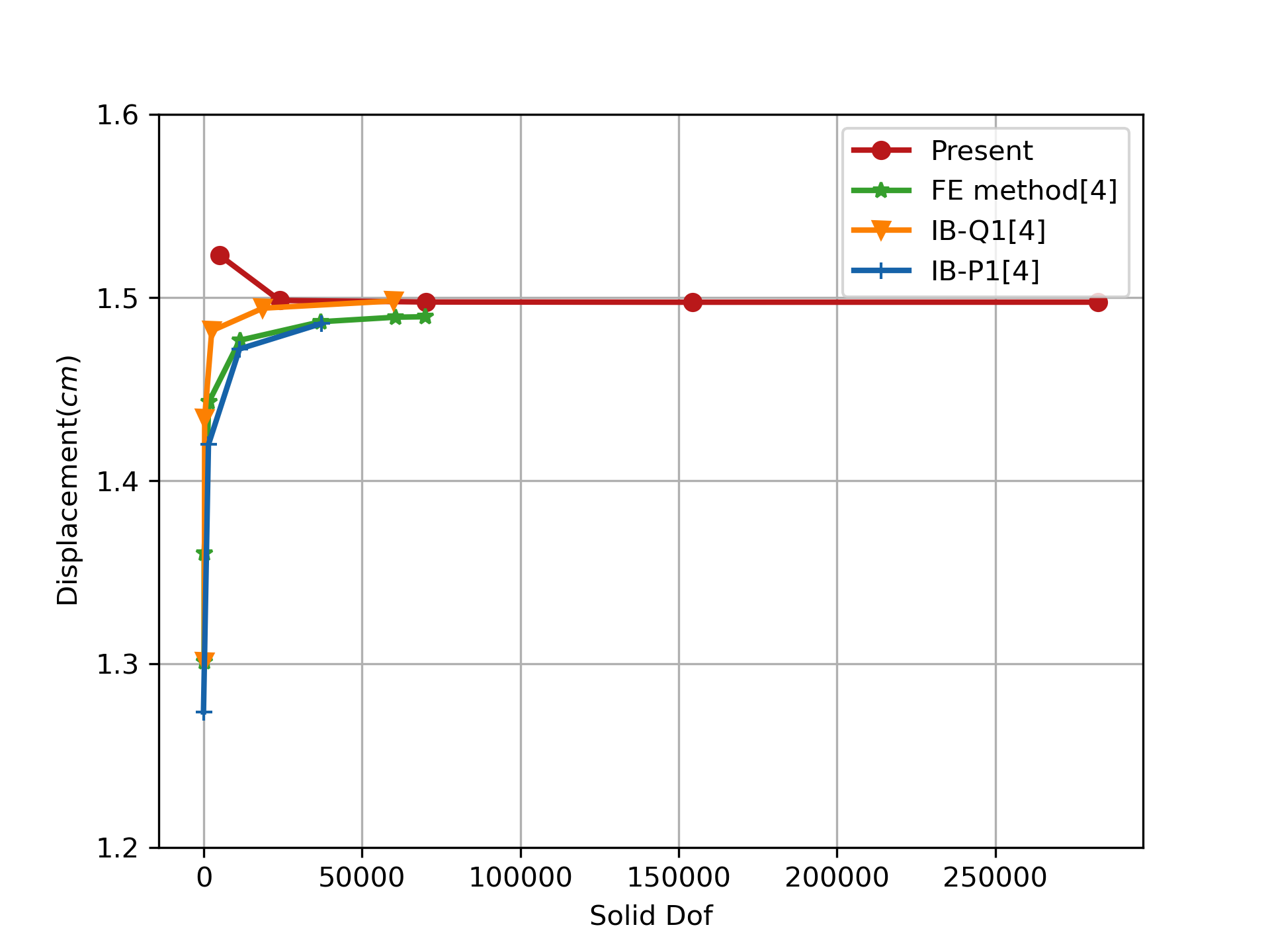}
        \subcaption{Difference numbers of Dof on the displacement of Point A for the Cook's membrane model.}
        \label{fig::Cook's membrane::displacement::nodes}
    \end{minipage}
    \caption{Corner $y$-displacement for the anisotropic Cook's membrane benchmark.  }
\end{figure}

To verify the accuracy of our computational results, we primarily focus on the deformation 
of the Cook's membrane, the determinant of the element Jacobian matrix $(J)$, 
and the vertical displacement at the upper right corner $A=(7.06~\text{cm}, 8.0~\text{cm}, 4.5~\text{cm})$ of the membrane. 
Fig.\,\ref{fig::cook_membrane::anisotropic::deformation} illustrates the deformation of the Cook membrane at the final time from different angles.
The results appear to be consistent with those in reference \cite{vadala2020stabilization}.
Fig.\,\ref{fig::cook_membrane::anisotropic::Jacobian} presents the computational results for $J$. Upon examining the figure, 
it is clear that $J$ is uniformly close to $1$ throughout the entire solid. Fig.\,\ref{fig::cook_membrane::anisotropic::volume_change} compares the volume change rates of solids calculated by our method under various degrees of freedom with those reported in reference \cite{vadala2020stabilization}. 
The analysis reveals that the volume change rates derived from our calculations range from $-0.00062\%$ to $-0.0088\%$, 
closely aligning with the data presented in reference \cite{vadala2020stabilization}. Of particular note, the results obtained using $P2$ elements 
in our study are nearly identical to those calculated using $Q1$ elements as documented in reference \cite{vadala2020stabilization}.
Fig.\,\ref{fig::Cook's membrane::displacement} illustrates the variation of the displacement of point $A$ along the y-axis over time.
Fig.\,\ref{fig::Cook's membrane::displacement::nodes} shows the variation in the maximum displacement of point $A$ along the $y$-axis with different Dofs of the solid. 
Furthermore, Fig.\,\ref{fig::Cook's membrane::displacement::nodes} demonstrates that the displacement of point $A$ in the $y$-axis, 
as calculated by our method, is essentially consistent with the results reported in the literature \cite{vadala2020stabilization}.
All of these results substantiate the accuracy of our algorithm.


\section{Cardiac Dynamics}
\subsection{Dynamic left ventricle model }

\begin{figure}[!ht]
\centering
\includegraphics[width=0.4\textwidth]{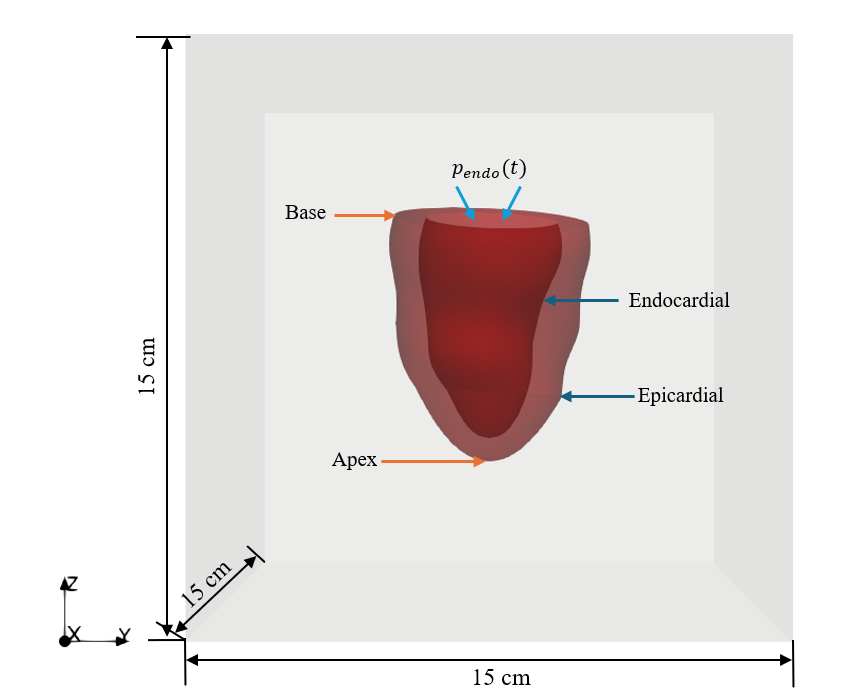}
\caption{Setup of setting up the LV model in the IB framework: The gray region represents the fixed computational domain $\Omega$, while the red region corresponds to the deformable left ventricle.}
\label{fig::left_ventricle::model::1}
\end{figure}

As one of the four chambers of the heart, the left ventricle (LV) plays a crucial role in the heart's pumping function. This section is based on a previously developed image-based geometric model \cite{gao2017changes}, similar to many existing studies \cite{gao2014quasi, gao2014dynamic, cai2015multi}, 
considers only a portion of the LV (see Fig.\,\ref{fig::left_ventricle::model::1}) to 
investigate a simplified cardiac mechanics problem relevant to physiological conditions.

In the LV model shown in Fig.\,\ref{fig::left_ventricle::model::1}, zero normal and circumferential displacements are applied at the base, 
allowing only radial expansion within the plane of the base. Meanwhile, the internal blood pressure is modeled by a uniform pressure applied to the endocardial surface, 
varying according to the prescribed waveform $p_{\text{endo}}(t)$ as referenced in \cite{thekkethil2023stabilized}. 
The maximum endocardial pressures during diastole and systole are $1067~\text{Pa}$ and $14530~\text{Pa}$, respectively, 
as defined below.
\begin{equation*}
    p_{\text{endo}}(t) =
\begin{cases}
\frac{1067 \, t}{0.2} \text{ Pa}, & \text{if } t < 0.2, \\[8pt]
1067 \text{ Pa}, & \text{if } 0.2 < t < 0.5, \\[8pt]
1067 + 13460 \left( 1 - \exp\left( -\frac{(t - 0.5)^2}{0.004} \right) \right) \text{ Pa}, & \text{if } 0.5 < t < 0.65, \\[8pt]
1067 + 13460 \left( 1 - \exp\left( -\frac{(0.8 - t)^2}{0.004} \right) \right) \text{ Pa}, & \text{if } 0.65 < t < 0.8.
\end{cases}
\end{equation*}
We model the left ventricular myocardium as a nonlinear, anisotropic hyperelastic material. To account for the effect of active tension, 
the first Piola-Kirchhoff stress tensor of the myocardium is modelled as the sum of the active and passive stress,
\begin{align*}
    \mathbb{P}^{e}=\frac{\partial\Psi}{\partial\mathbb{F}}+\mathbb{P}^{a}.
\end{align*} 
When the passive stress component is described using the HO model, the active stress component $\mathbb{P}^{a}$ is given by:
\begin{align*}
    \mathbb{S}_{a}=T(t,I_{4f})\mathbf{e}_{f}\otimes\mathbf{e}_{f},
\end{align*}
where the scalar tension $T=T(t,I_{4f})$ is a function of time $t$ and $I_{4f}$.
Specifically, 
\begin{align*}
    T(t,I_{4f})=T_{a}(t)\left(1+4.9(\sqrt{I_{4f}}-1)\right),
\end{align*}
where $T_{a}(t)$ is the active tension, which is determined empirically. 
Based on the maximum active tension value of a healthy LV reported in \cite{gao2014dynamic}, we set its expression according to \cite{thekkethil2023stabilized}, as follows:
\begin{equation*}
    T_a(t) =
    \begin{cases} 
    0 \text{ Pa}, & \text{if } t < 0.5, \\ 
    84260 \left( 1 - \exp \left( \frac{-(t - 0.5)^2}{0.005} \right) \right) \text{ Pa}, & \text{if } 0.5 < t < 0.65, \\ 
    84260 \left( 1 - \exp \left( \frac{-(0.8 - t)^2}{0.005} \right) \right) \text{ Pa}, & \text{if } 0.65 < t < 0.8.
    \end{cases} 
\end{equation*}
The detailed physical and numerical parameters are listed in Table \ref{tab::left::ventricle::params}.

\begin{figure}[!ht]
    \centering
    \begin{minipage}{0.45\textwidth}
        \centering
        \includegraphics[width=0.55\textwidth]{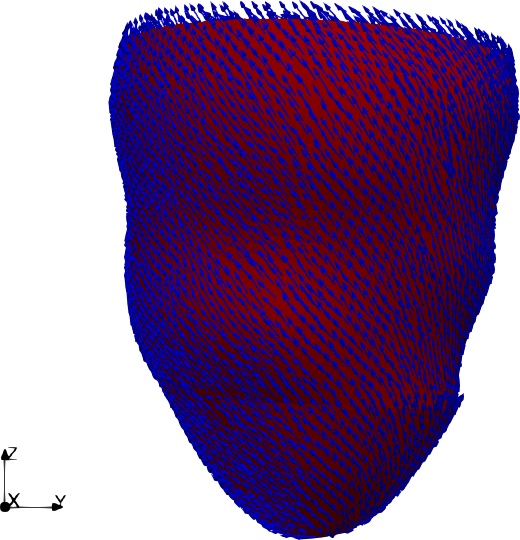}
        \subcaption{Fiber: $\mathbf{e}_{f}$}
        \label{fig::left_ventricle::f0}
    \end{minipage}
    \begin{minipage}{0.45\textwidth}
        \centering
        \includegraphics[width=0.55\textwidth]{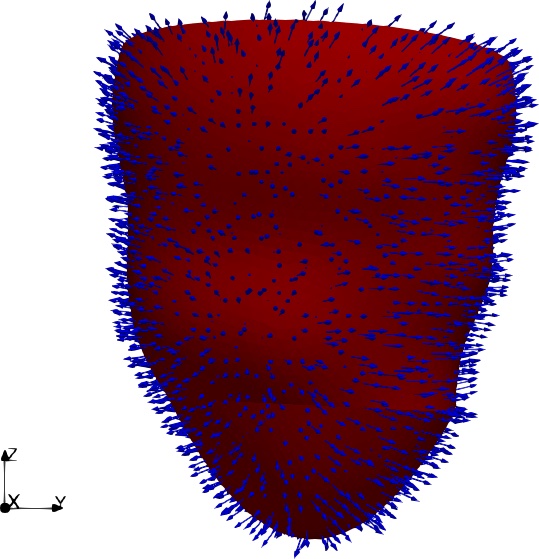}
        \subcaption{Fiber: $\mathbf{e}_{s}$}
        \label{fig::left_ventricle::s0}
    \end{minipage}
    \caption{Fiber distribution in the LV model. Blue arrows indicate the direction of fibers, and red represents the LV model.}
\end{figure}

\begin{table}[!h]
    \centering
    \begin{tabular}{lcc}
        \toprule
        \textbf{Symbol} & \textbf{Value} & \textbf{Unit}  \\
        \midrule
        $\rho$  & 1.0   & $\frac{\text{g}}{\text{cm}^3}$ \\
        $T$ & 2    & $\text{s}$  \\
        $\mu$   & 0.04  & $\frac{\text{dyn}\cdot \text{s}}{\text{cm}^{2}}$\\
        \midrule
        $a$  & 2244.87  & $\frac{\text{dyn}}{\text{cm}^2}$\\
        $a_{f}$  & 24267 & $\frac{\text{dyn}}{\text{cm}^2}$ \\
        $a_{s}$  & 5562.38 & $\frac{\text{dyn}}{\text{cm}^2}$ \\
        $a_{fs}$  & 3905.16 & $\frac{\text{dyn}}{\text{cm}^2}$ \\
        $b$  & 1.6215 &  -    \\
        $b_{f}$  & 1.8268 & -    \\
        $b_{s}$  &  0.7746 & -   \\
        $b_{fs}$  & 1.695 & - \\
        $\kappa_\text{s}$ &  $10^6$ & - \\
        \bottomrule
    \end{tabular}
    \caption{Parameters for the dynamic left ventricle model.}
    \label{tab::left::ventricle::params}
\end{table}

\begin{figure}[!ht]
    \centering
    \begin{minipage}{0.32\textwidth}
        \centering
        \includegraphics[width=0.7\textwidth]{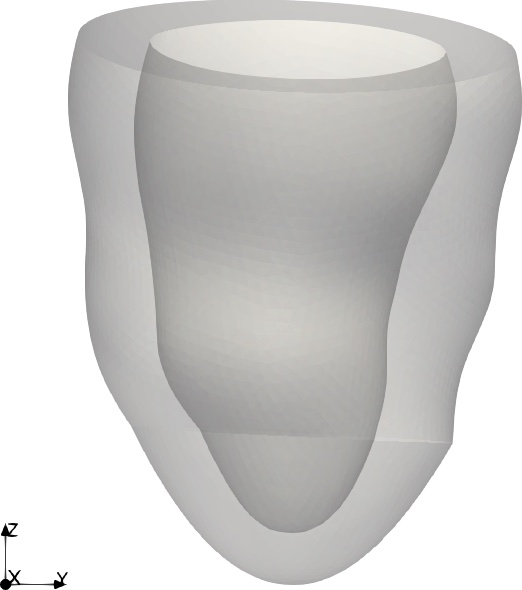}
        \subcaption{Original position}
        \label{fig::left_ventricle::deformation::origin}
    \end{minipage}
    \begin{minipage}{0.32\textwidth}
        \centering
        \includegraphics[width=0.65\textwidth]{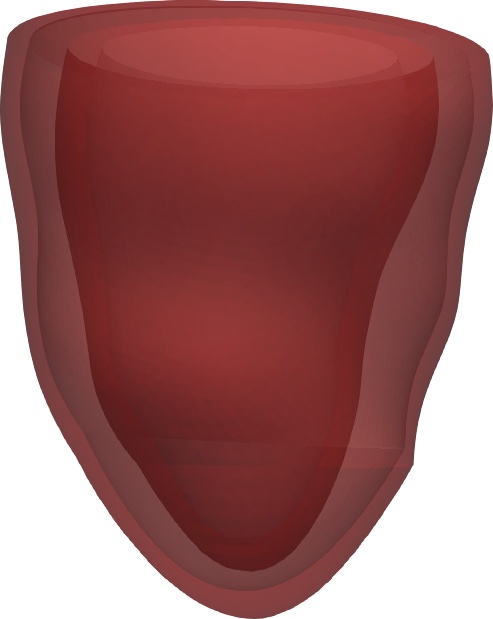}
        \subcaption{Diastole}
        \label{fig::left_ventricle::deformation::diastole}
    \end{minipage}
    \begin{minipage}{0.32\textwidth}
        \centering
        \includegraphics[width=0.60\textwidth]{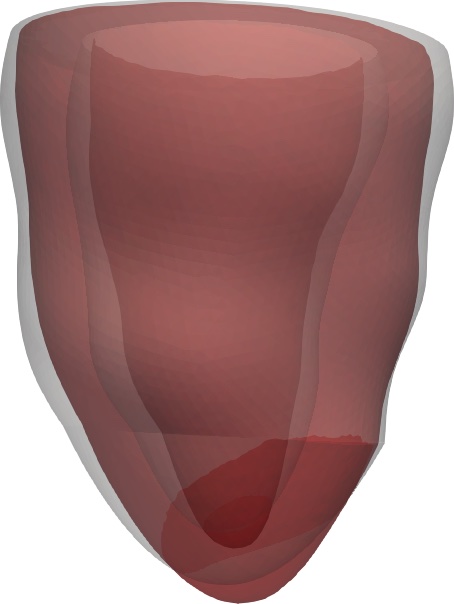}
        \subcaption{Systole}
        \label{fig::left_ventricle::deformation::systole}
    \end{minipage}
    \caption{Schematic illustration of left ventricular deformation. The red color indicates the deformed position, while the grey 
    color indicates the original position.}
    \label{fig::left_ventricle::deformation}
\end{figure}

\begin{figure}[!ht]
    \centering
    \begin{minipage}{0.32\textwidth}
        \centering
        \includegraphics[width=1.0\textwidth]{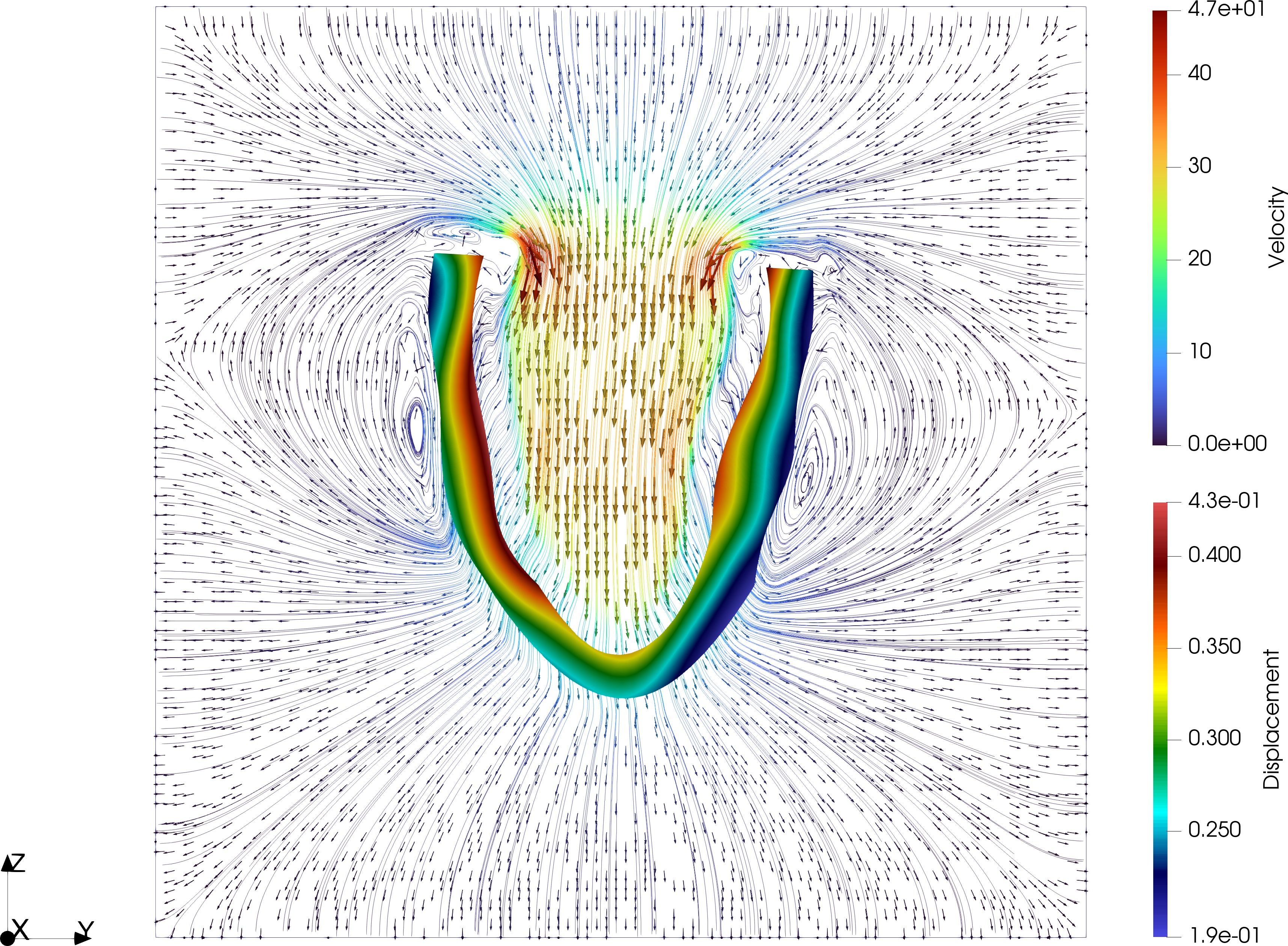}
        \subcaption{$0.12~\text{s}$}
        \label{fig::left_ventricle::0.12s}
    \end{minipage}
    \begin{minipage}{0.32\textwidth}
        \centering
        \includegraphics[width=1.0\textwidth]{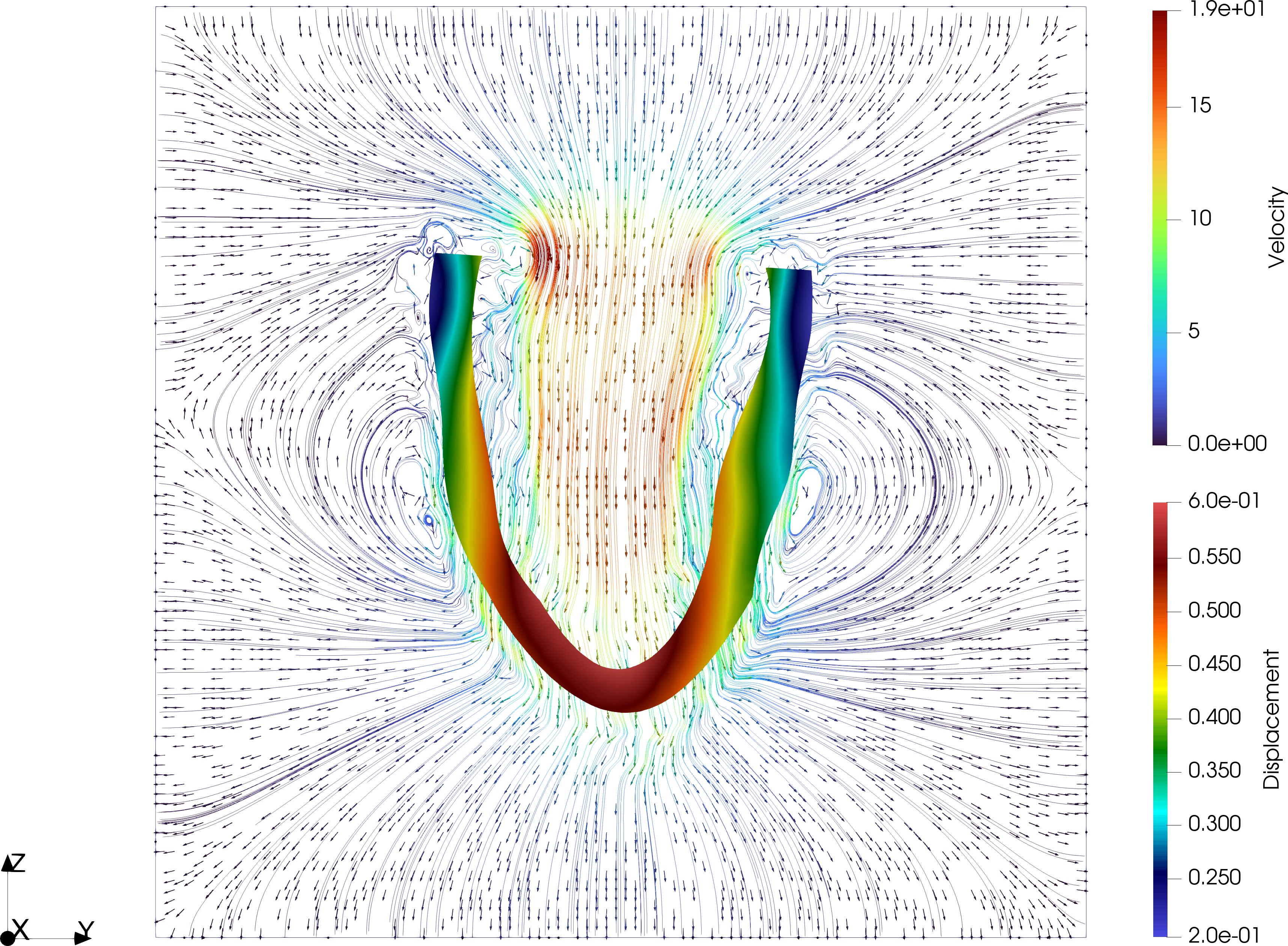}
        \subcaption{$0.45~\text{s}$}
        \label{fig::left_ventricle::0.45s}
    \end{minipage}
    \begin{minipage}{0.32\textwidth}
        \centering
        \includegraphics[width=1.0\textwidth]{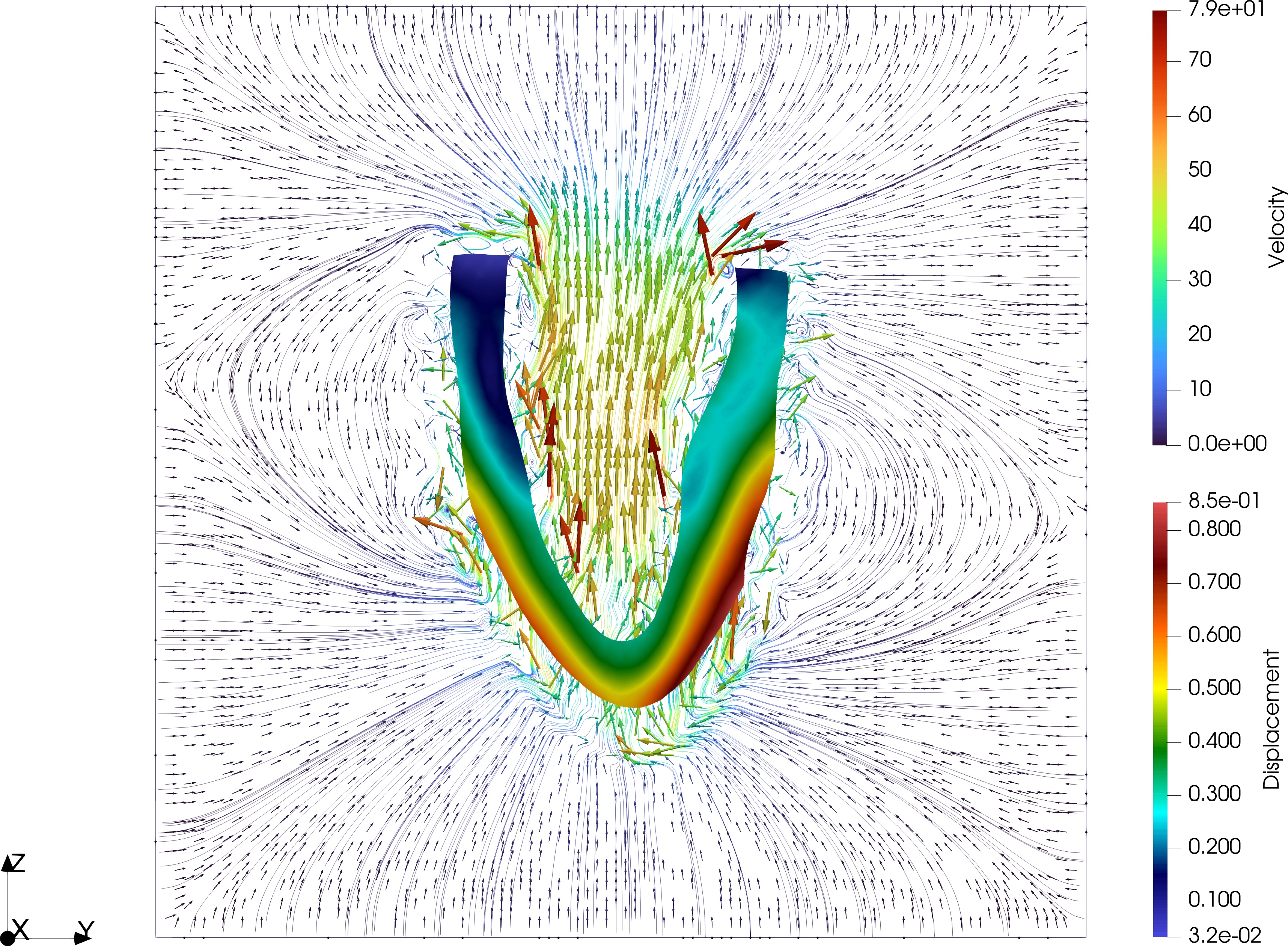}
        \subcaption{$0.72~\text{s}$}
        \label{fig::left_ventricle::0.72s}
    \end{minipage}
    \caption{Deformation of a left ventricular slice and fluid velocity streamlines at different times.}
    \label{fig::left_ventricle::slice}
\end{figure}

\begin{figure}[!ht]
    \centering
    \begin{minipage}{0.45\textwidth}
        \centering
        \includegraphics[width=1.0\textwidth]{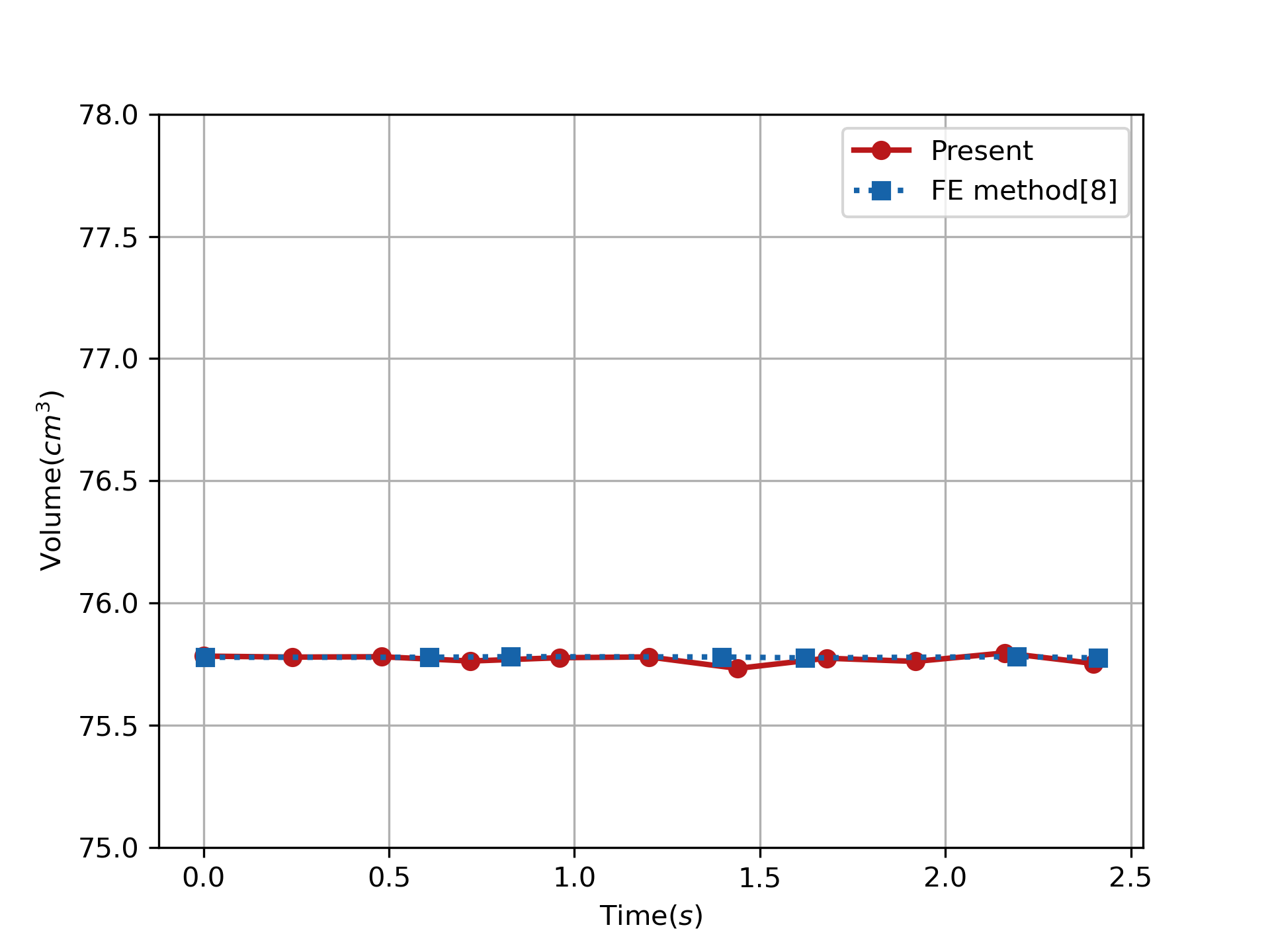}
        \subcaption{Volume}
        \label{fig::left_ventricle::volume1}
    \end{minipage}
    \begin{minipage}{0.45\textwidth}
        \centering
        \includegraphics[width=1.0\textwidth]{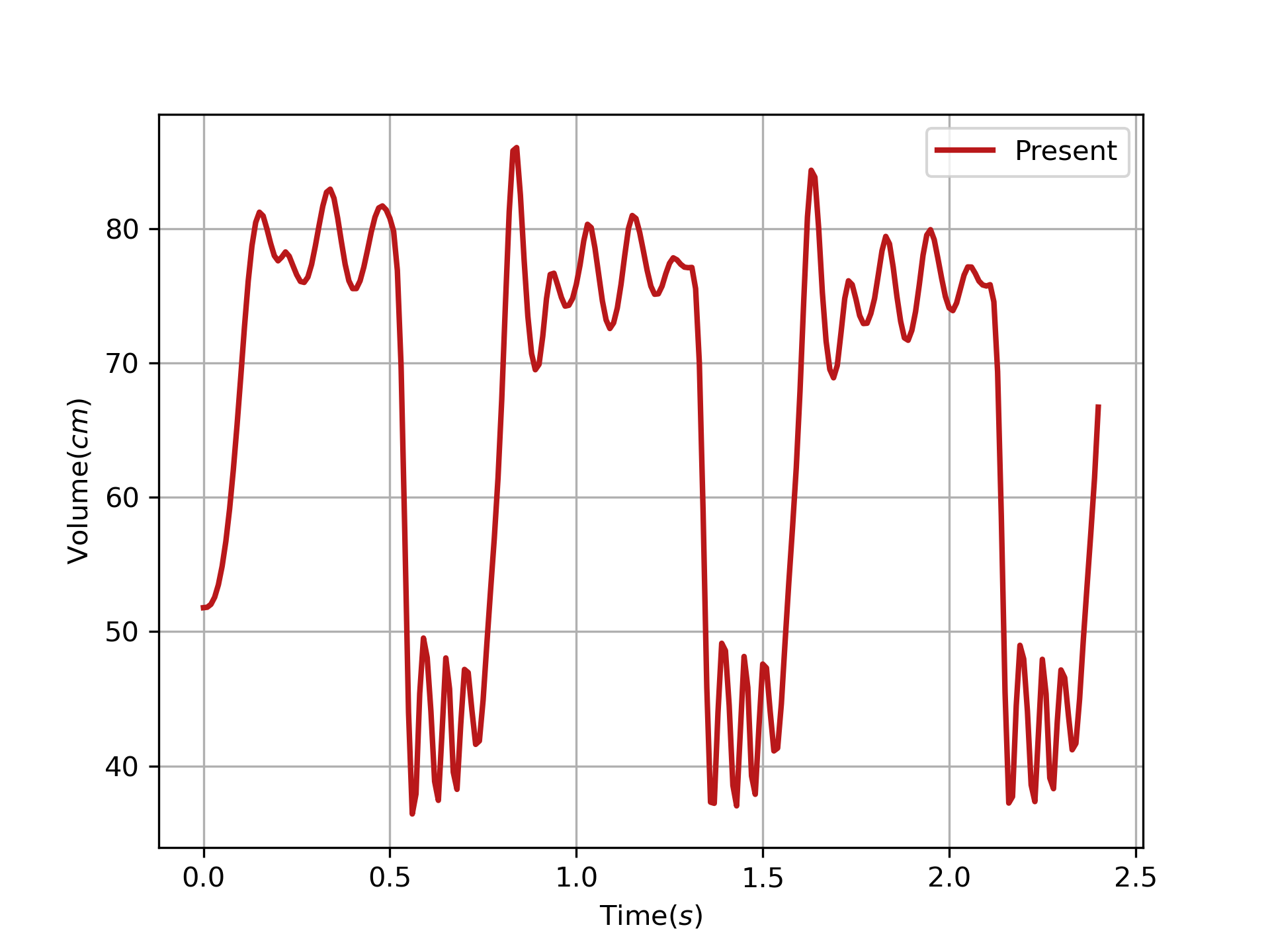}
        \subcaption{Capacity}
        \label{fig::left_ventricle::volume2}
    \end{minipage}
    \caption{Time-dependent changes in left ventricular volume and ventricular capacity. }
\end{figure}

For this model, we simulated three consecutive cardiac cycles, corresponding to the diastolic and systolic phases. 
Fig.\,\ref{fig::left_ventricle::deformation} illustrates the deformation of the left ventricle, 
where Fig.\,\ref{fig::left_ventricle::deformation::diastole} shows the left ventricle in diastole, and Fig.\,\ref{fig::left_ventricle::deformation::systole} 
shows the left ventricle in systole. 
Fig.\,\ref{fig::left_ventricle::slice} shows the deformation of a left ventricular slice and the velocity streamlines at $0.12~\text{s}$, $0.45~\text{s}$, and $0.72~\text{s}$. 
From the Fig.\,\ref{fig::left_ventricle::slice}, it can be seen that at $0.12~\text{s}$ and $0.45~\text{s}$, the LV is in the diastolic phase, 
during which blood flows into the LV. At $0.45~\text{s}$, the LV reaches near-maximal diastole, 
and the velocity streamline plot shows a reduced flow velocity, 
indicating that the blood inflow rate decreases and the left ventricular volume approaches its maximum. 
At $0.72~\text{s}$, the LV is in the systolic phase, during which blood is ejected from the LV. This process is consistent with 
the physiological motion of the LV during cardiac pumping.
Fig.\,\ref{fig::left_ventricle::volume1} illustrates the temporal variation of left ventricular 
volume. Given the incompressibility of myocardial tissue, the left ventricular volume 
calculated by the current method is consistent with that derived from a purely solid model, 
remaining nearly constant throughout the cardiac cycle. Fig.\,\ref{fig::left_ventricle::volume2}, in contrast, 
depicts the changes in left ventricular volume over time.

\begin{table}[!h]
    \centering
    \caption{Profiling results that identify the hot-spots in the present IB method.}
    \begin{tabular}{lcc|cc}
        \toprule
        & \multicolumn{2}{c|}{\textbf{CPU-GPU Heterogeneous}} & \multicolumn{2}{c}{\textbf{Full GPU (Present)}} \\
        \cmidrule(lr){2-3} \cmidrule(lr){4-5}
        & (\%) & (s) & (\%) & (s) \\
        \midrule
        Fluid Solver (Poisson equation) & 71.35 & 10827.8 & 67.6 & 50.1 \\
        Fluid Solver (Helmholtz equation) & 26.99 & 4095.58 & 31 & 22.8 \\
        Fluid Solver (Convective Term) & 0.34 & 51.59 & 0 & 0.9 \\
        Interpolation & 1.25 & 189.31 & 1.4 & 0.93 \\
        Spreading & 1.25 & 189.31 & 1.4 & 0.93 \\
        Solid Solver & 1.25 & 189.31 & 1.4 & 0.93 \\ 
        Solid boundary  & & & & 1.63 \\
        \bottomrule
    \end{tabular}
\end{table}

\subsection{Aortic valve model }

The aortic valve is located between the left ventricle and the aorta, controlling blood flow into the aorta and preventing 
blood from flowing back, thus ensuring normal heart function. This section will delve into the physiological behavior of the aortic valve. 
The aortic valve typically consists of three crescent-shaped leaflets, which exhibit highly anisotropic, incompressible, hyperelastic, 
and nonlinear mechanical properties. To effectively describe these mechanical characteristics, 
we have adopted Fiber-reinforced hyperelastic model 2 to represent the mechanical behavior of the aortic valve leaflets.
The aortic valve model used in this study is based on the porcine heart valve. 
The thickness of the valve leaflets is $0.04~\text{cm}$, and the length of the outer rigid circular tube surrounding the aortic valve leaflets is $13~\text{cm}$, 
with a wall thickness of $0.15~\text{cm}$ and an inner radius of $1.3~\text{cm}$ \cite{cai2021comparison, zhu2017vitro, flamini2016immersed}. 
The complete FSI model is shown in Fig.\,\ref{fig::aortic_valve::model::1}. 
In the numerical calculation, the solid model is discretized into 229,865 tetrahedral elements and 7669 nodes, 
while the fluid domain is discretized into $80\times 80 \times 128$ Cartesian grid elements. The relevant physical and numerical parameters is provided in Table \ref{tab::aortic::valve::params}.
\begin{figure}[!ht]
    \centering
    \includegraphics[width=1.0\textwidth]{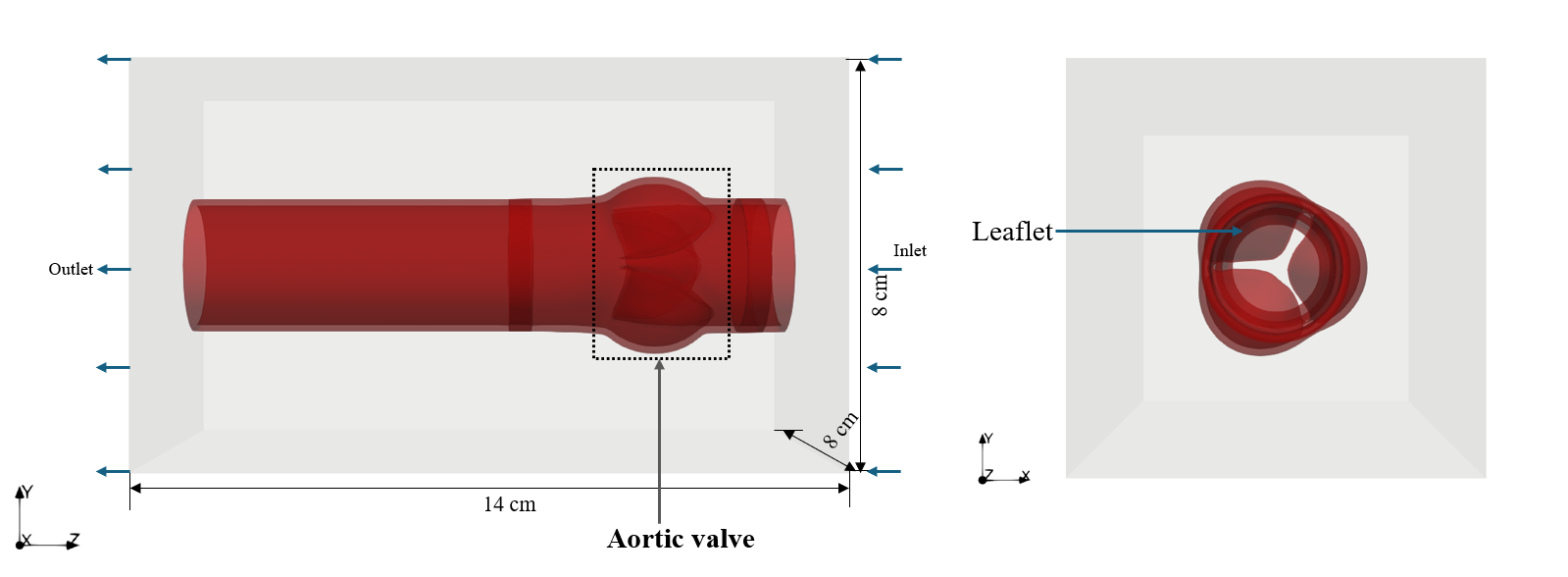}
    \caption{Setup of setting up the AV model in the IB framework: the gray region represents the fixed computational domain $\Omega$, while the red region corresponds to the deformable aortic valve.}
    \label{fig::aortic_valve::model::1}
\end{figure}

\begin{figure}[!ht]
    \centering
    \includegraphics[width=0.3\textwidth]{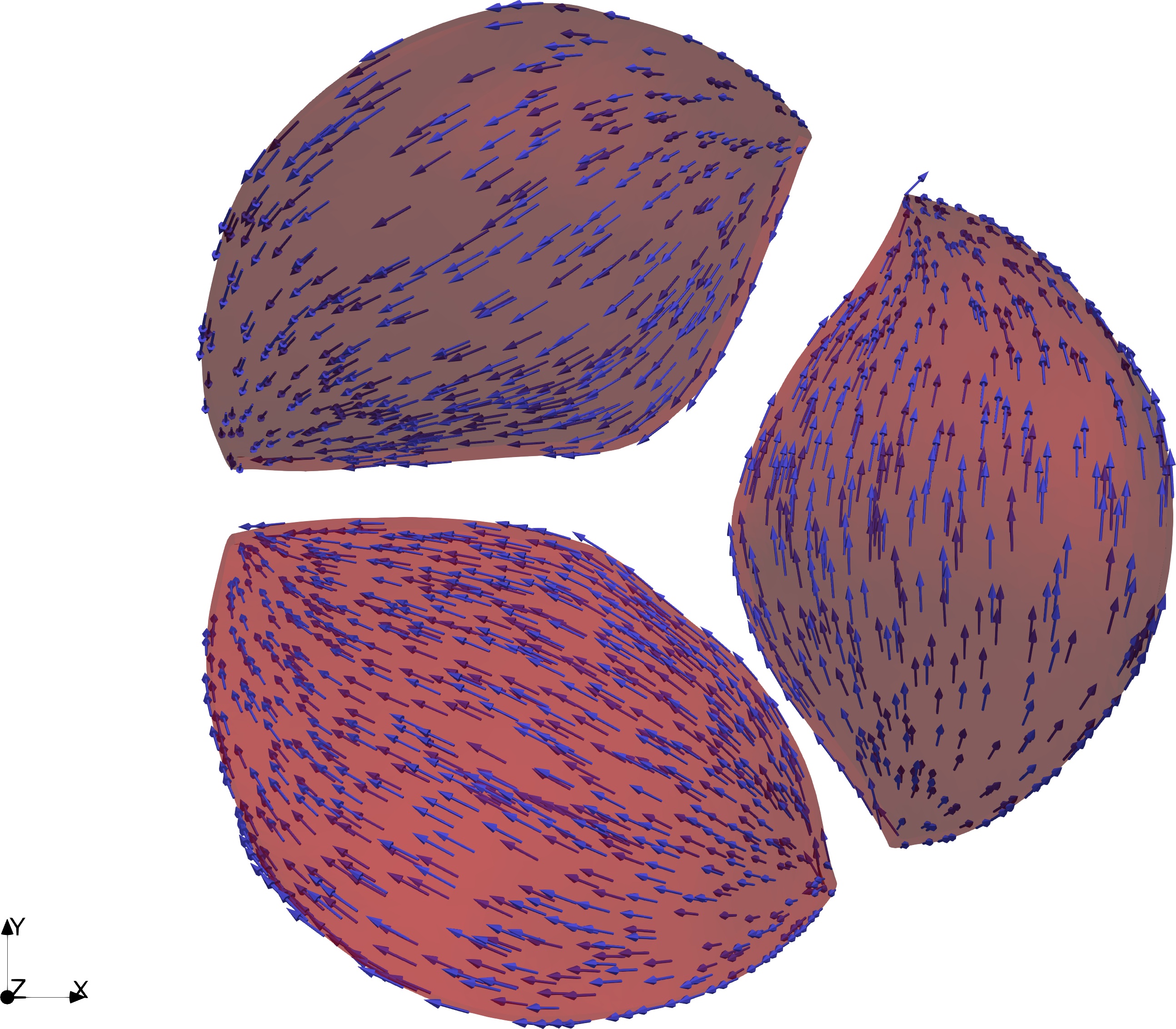}
    \caption{Fiber distribution in the AV model. Blue arrows indicate the direction of fibers, and red represents the AV leaflets.}
    \label{fig::aortic_valve::model::1}
\end{figure}

\begin{table}[!h]
    \centering
    \begin{tabular}{lcc}
        \toprule
        \textbf{Symbol} & \textbf{Value} & \textbf{Unit}  \\
        \midrule
        $\rho$  & 1.0   & $\frac{\text{g}}{\text{ml}}$ \\
        $T$ & 1.635    & $\text{s}$  \\
        $\mu$   & 4  & $\text{cP}$\\
        \midrule
        $C_{10}$  & 1210  & $\text{Pa}$\\
        $C_{01}$  & 7.99 & - \\
        $a_{f}$  & 24230 & $\text{Pa}$ \\
        $b_{f}$  & 57.62 & -    \\
        $\kappa_\text{s}$ &  $10^6$ & - \\
        \bottomrule
    \end{tabular}
    \caption{Parameters for the AV model.}
    \label{tab::aortic::valve::params}
\end{table}

\begin{figure}[!ht]
    \centering
    \begin{minipage}{0.45\textwidth}
        \centering
        \includegraphics[width=0.8\textwidth]{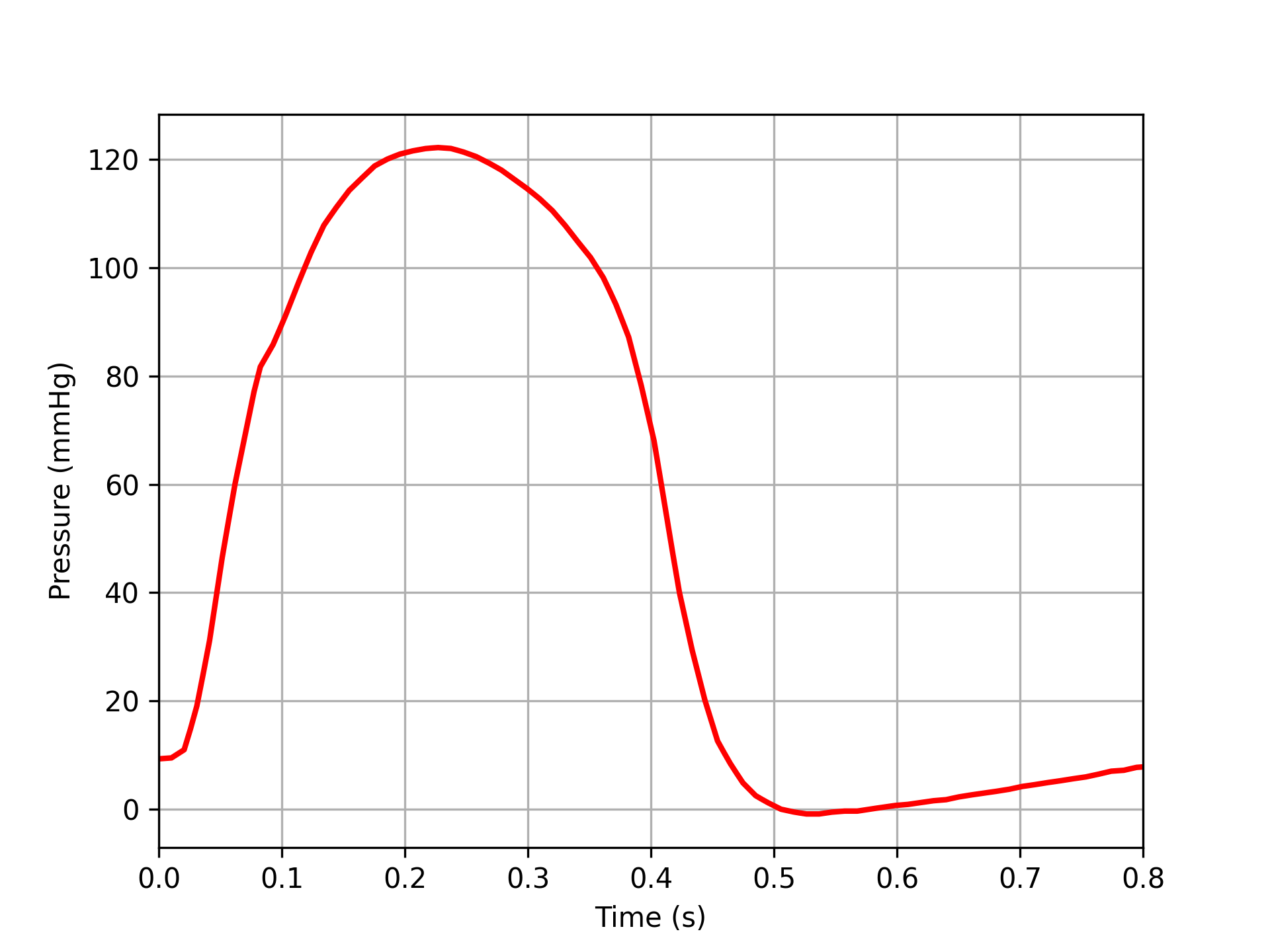}
        \subcaption{Left ventricular pressure curve at the inlet. }
        \label{fig::aortic_valve::inlet pressure}
    \end{minipage}
    \begin{minipage}{0.45\textwidth}
        \centering
        \includegraphics[width=1\textwidth]{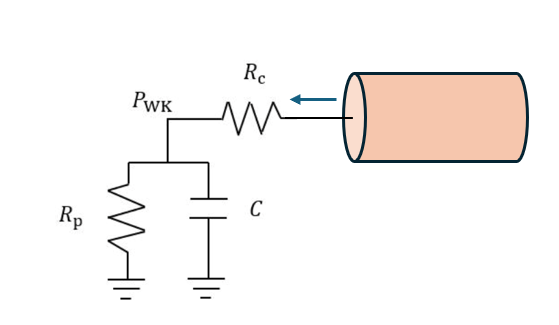}
        \subcaption{Windkessel model at the outlet}
        \label{fig::aortic_valve::windkessel}
    \end{minipage}
    \caption{Boundary conditions applied within the AV IB framework. }
\end{figure}

For this problem, the right end of the outer tube is the inlet, corresponding to the left ventricle, and the left end is the outlet, corresponding to the aorta. The relevant boundary conditions are as follows:  
\begin{itemize}
    \item   The physiological pressure of the left ventricle is applied at the inlet of the fluid region to drive blood flow through the aortic valve, with the left ventricular pressure curve shown over one cardiac cycle in Fig.\,\ref{fig::aortic_valve::inlet pressure}. 
    \item   The outlet of the fluid region uses a three-element Windkessel model to apply dynamic pressure loading, as shown in Fig.\,\ref{fig::aortic_valve::windkessel}, 
            driving the aortic valve model. The parameters of the Windkessel model include characteristic resistance $R_{c}=0.033~\text{mmHgml}^{-1}$, 
            peripheral resistance $R_{p}=0.79~\text{mmHg}\!\cdot\!{ml}^{-1}$, arterial compliance $C=90~\text{ml}\!\cdot\!\text{mmHg}^{-1}$, with an initial pressure of $P_{Wk}=85~\text{mmHg}$, and $P_{Ao}=P_{Wk}$, 
            while the outlet pressure is set to 0. 
            These parameter values are referenced from \cite{cai2021comparison, stergiopulos1999total}.
    \item Zero pressure boundary conditions are applied to the remaining boundaries of the fluid region.
    \item The outer tube of the aortic valve is fixed, and a penalty parameter of $\kappa = 5\times 10^{6} $ is applied.
\end{itemize}


\subsection{Mitral valve model}
The mitral valve is located between the left atrium and the left ventricle, controlling blood flow into the left ventricle and preventing backflow, thereby ensuring normal blood circulation and reducing the burden on the heart. 
This section will examine the physiological behavior of the mitral valve.
The mitral valve has a complex tissue structure, including two main valve leaflets, the mitral annulus, chordae tendineae, 
and papillary muscles. The valve leaflets are asymmetrical in structure, and the mitral annulus is fixed to the left ventricular wall. 
The interaction between the leaflets and the annulus prevents blood from flowing back into the left atrium during cardiac contraction.
The chordae tendineae connect the valve leaflets to the papillary muscles, primarily controlling the motion of the leaflets, 
while the papillary muscles, fixed to the left ventricular wall, serve as anchor points for the chordae tendineae, 
regulating the motion of the valve leaflets.
To describe the highly anisotropic, incompressible, hyperelastic, and nonlinear mechanical properties of the mitral valve leaflets, 
we use the Fiber-reinforced hyperelastic model 1 to characterize their constitutive relations \cite{cai2019some, gao2014finite}. The mechanical properties of the chordae tendineae are described using 
a linear elastic Neo-Hookean model \cite{cai2019some, gao2014finite}:
$$\Psi=a(\bar{I}_{1}-3)+\frac{\kappa_{\text{stab}}}{2}(\text{ln}(J))^{2},$$
and its first Piola-Kirchhoff stress tensor is:
$$\mathbb{P}=2aJ^{-2/3}(\mathbb{F}-\frac{I_{1}}{3}\mathbf{F}^{-T})+\kappa_{\text{stab}}\text{ln}{J}\mathbb{F}^{-T}.$$
The thickness of the valve leaflets is $0.1~\text{cm}$, and the length of the outer rigid circular tube surrounding the mitral valve structure is $16~\text{cm}$, with a radius of $3.8~\text{cm}$.
The complete fluid-structure interaction (FSI) model is shown in Fig.\,\ref{fig::mitral_valve::model::1}.
In the numerical calculation, the solid model is discreted into 734645 tetrahedral elements and 188557 nodes, while the fluid domain is discreted into $80\times80\times128$ Cartesian grid elements.
The relevant physical and numerical parameters is provided in Table 4.
\begin{figure}[!ht]
    \centering
    \includegraphics[width=0.9\textwidth]{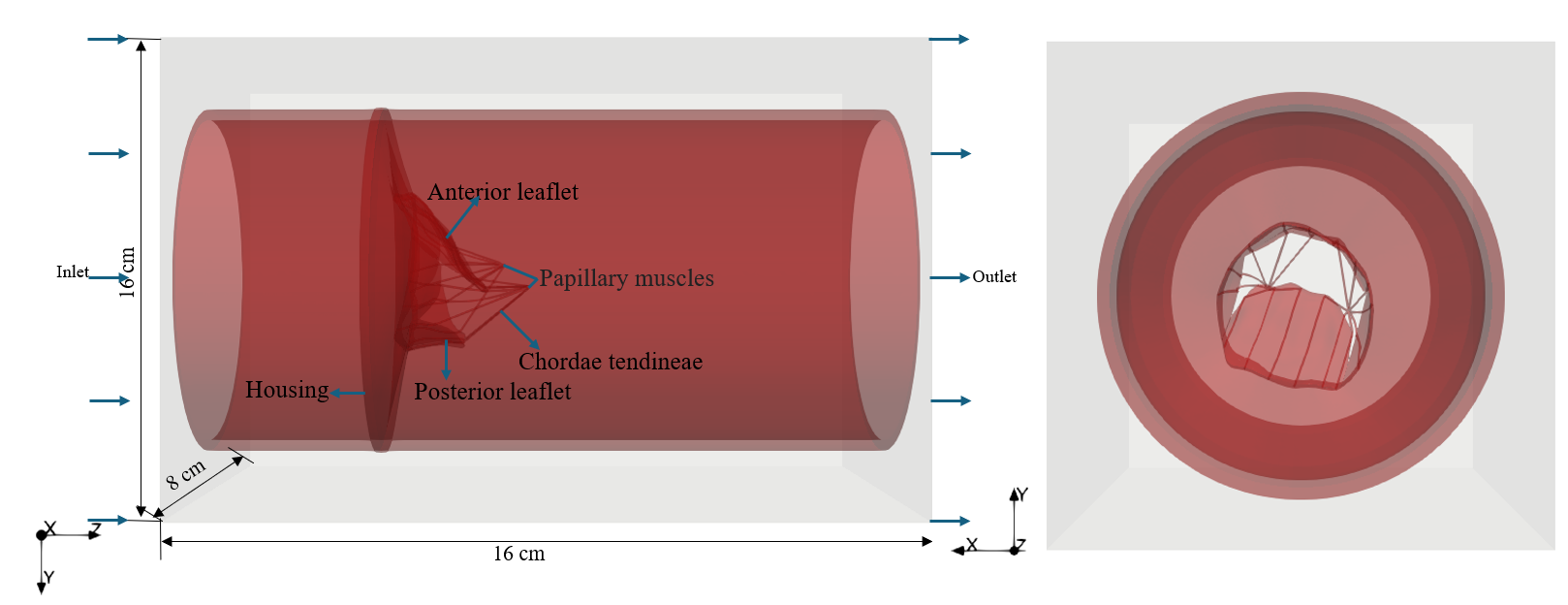}
    \caption{Setup of setting up the MV model in the IB framework: the gray region represents the fixed computational domain $\Omega$, while the red region corresponds to the deformable MV.}
    \label{fig::mitral_valve::model::1}
\end{figure}

\begin{figure}[!ht]
    \centering
    \includegraphics[width=0.3\textwidth]{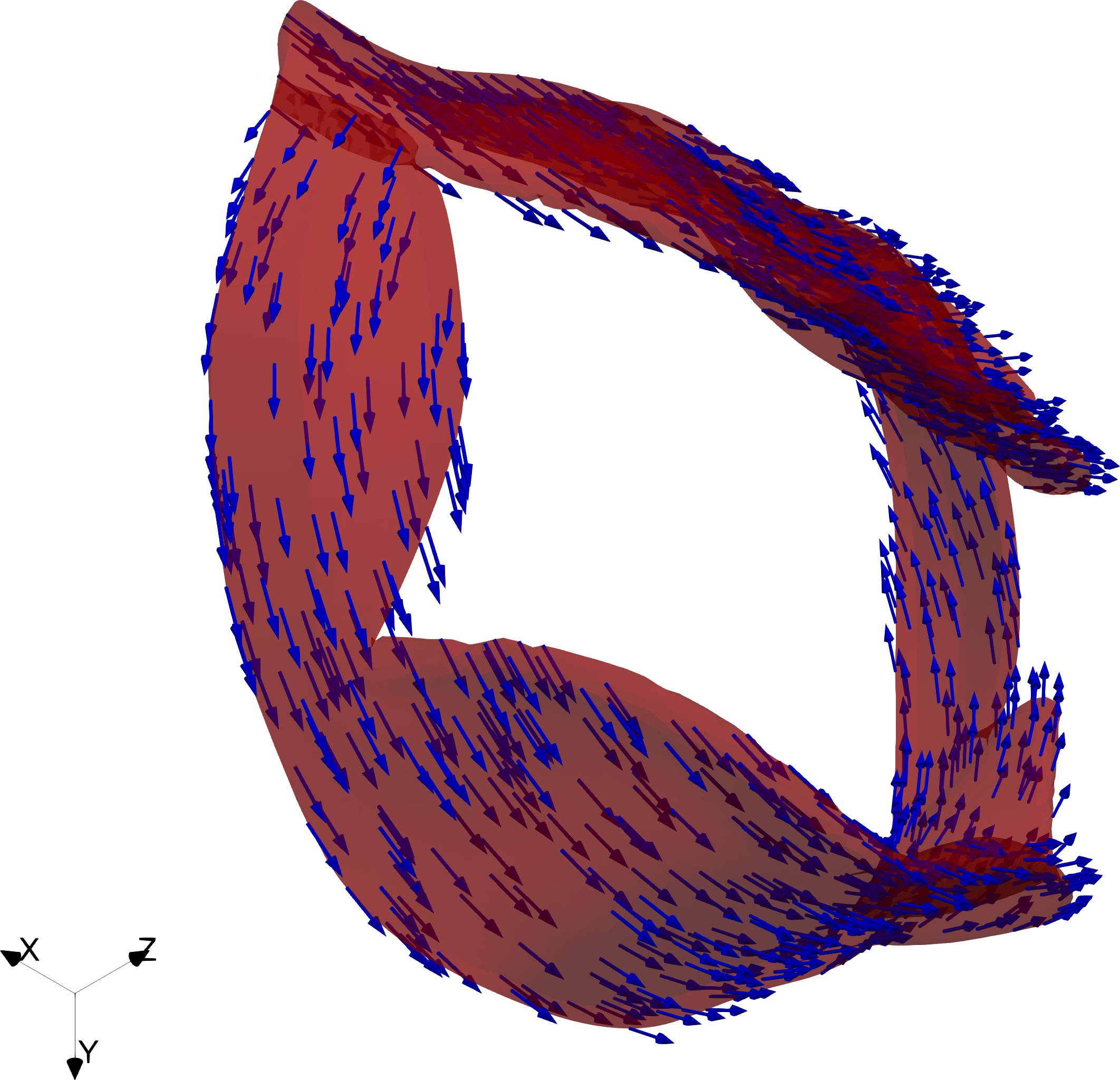}
    \caption{Fiber distribution in the MV model. Blue arrows indicate the direction of fibers, and red represents the MV leaflets.}
    \label{fig::mitral_valve::model::1}
\end{figure}

\begin{table}[!h]
    \centering
    \begin{tabular}{lccc}
        \toprule
        \textbf{Type} &\textbf{Symbol} & \textbf{Value} & \textbf{Unit}  \\
        \midrule
        -    &$\rho$  & 1.0   & $\frac{\text{g}}{\text{cm}^3}$ \\
        -    &$T$      & 2     & $\text{s}$  \\
        -    &$\mu$    & 0.04  & $\frac{\text{dyn}\cdot \text{s}}{\text{cm}^{2}}$\\
        \midrule
        Anterior leaflet    &$C_1$  & 17400  & $\text{Pa}$\\
        Anterior leaflet   &$a_{f}$  & 31300 & $\text{Pa}$ \\
        Anterior leaflet   &$b_{f}$  & 55.93 & -    \\
        -                           &$\kappa_\text{s}$ &  $10^6$ & - \\
        \midrule
        Posterior leaflet    &$C_1$  & 10200  & $\text{Pa}$\\
        Posterior leaflet   &$a_{f}$  & 50000 & $\text{Pa}$ \\
        Posterior leaflet    &$b_{f}$  & 63.48 & -    \\
        -                           &$\kappa_\text{s}$ &  $10^6$ & - \\
        \midrule
        Chordae tendineae(systole)  &   $C$   &9.0  &$\text{MPa}$ \\
        Chordae tendineae(diastole) &   $C$   &0.54  &$\text{MPa}$\\
        -                           &$\kappa_\text{s}$ &  $10^6$ & - \\      
        \bottomrule
    \end{tabular}
    \caption{Parameters for the mitral valve model.}
    \label{tab::mitral::valve::params}
\end{table}

The boundary conditions for the FSI system are set as follows: 
\begin{itemize}
    \item A pressure difference is defined between the inlet and outlet of the tube, with the transvalvular pressure difference based on a typical human physiological pressure curve (as shown in Figure 23).
    \item Zero-pressure boundary conditions are applied along the remainder boundaries of the fluid region \cite{gao2014finite, ma2013image}.
\end{itemize}
\begin{figure}[!ht]
    \centering
    \includegraphics[width=0.5\textwidth]{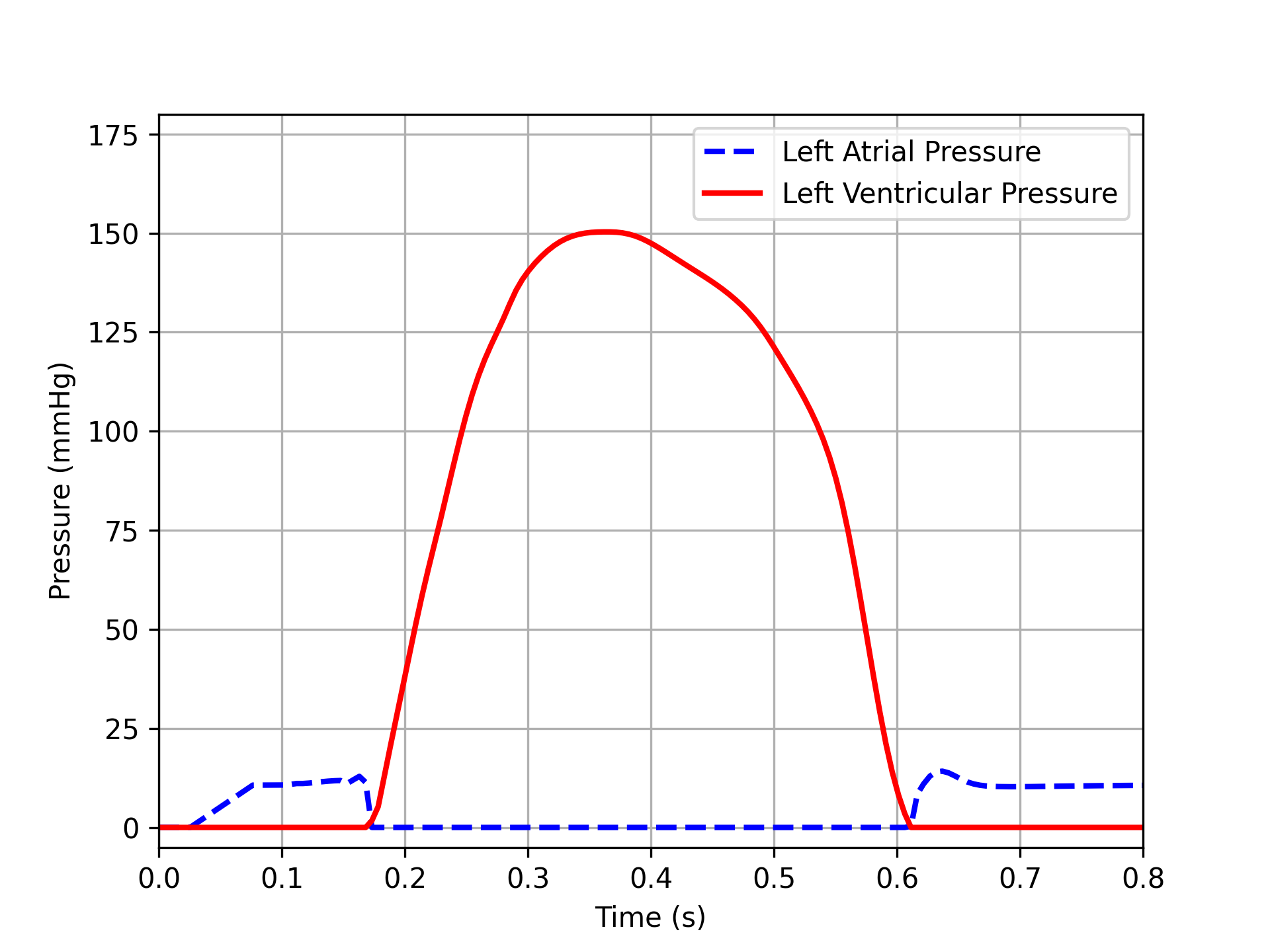}
    \caption{Setup of setting up the MV model in the IB framework: the gray region represents the fixed computational domain $\Omega$, while the red region corresponds to the deformable mitral valve.}
    \label{fig::mitral_valve::model::1}
\end{figure}

\begin{figure}[!ht]
    \centering
    \begin{minipage}[t]{0.48\textwidth}
        \centering
        \begin{minipage}[t]{\textwidth}
            \centering
            \includegraphics[width=1.0\textwidth]{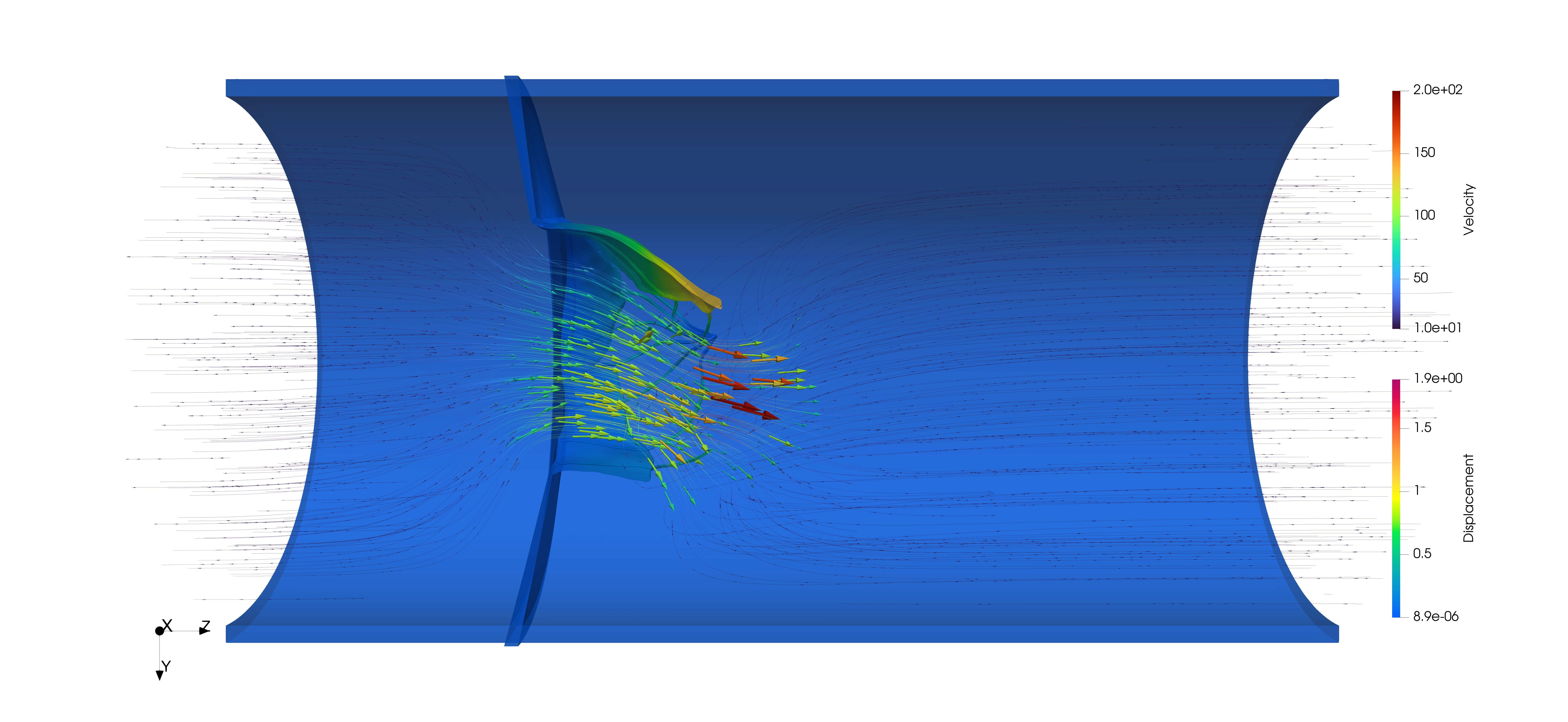}
            \subcaption{$0.10~\text{s}$}
            \label{fig::left_ventricle::a}
        \end{minipage}
        \vfill
        \begin{minipage}[t]{\textwidth}
            \centering
            \includegraphics[width=1.0\textwidth]{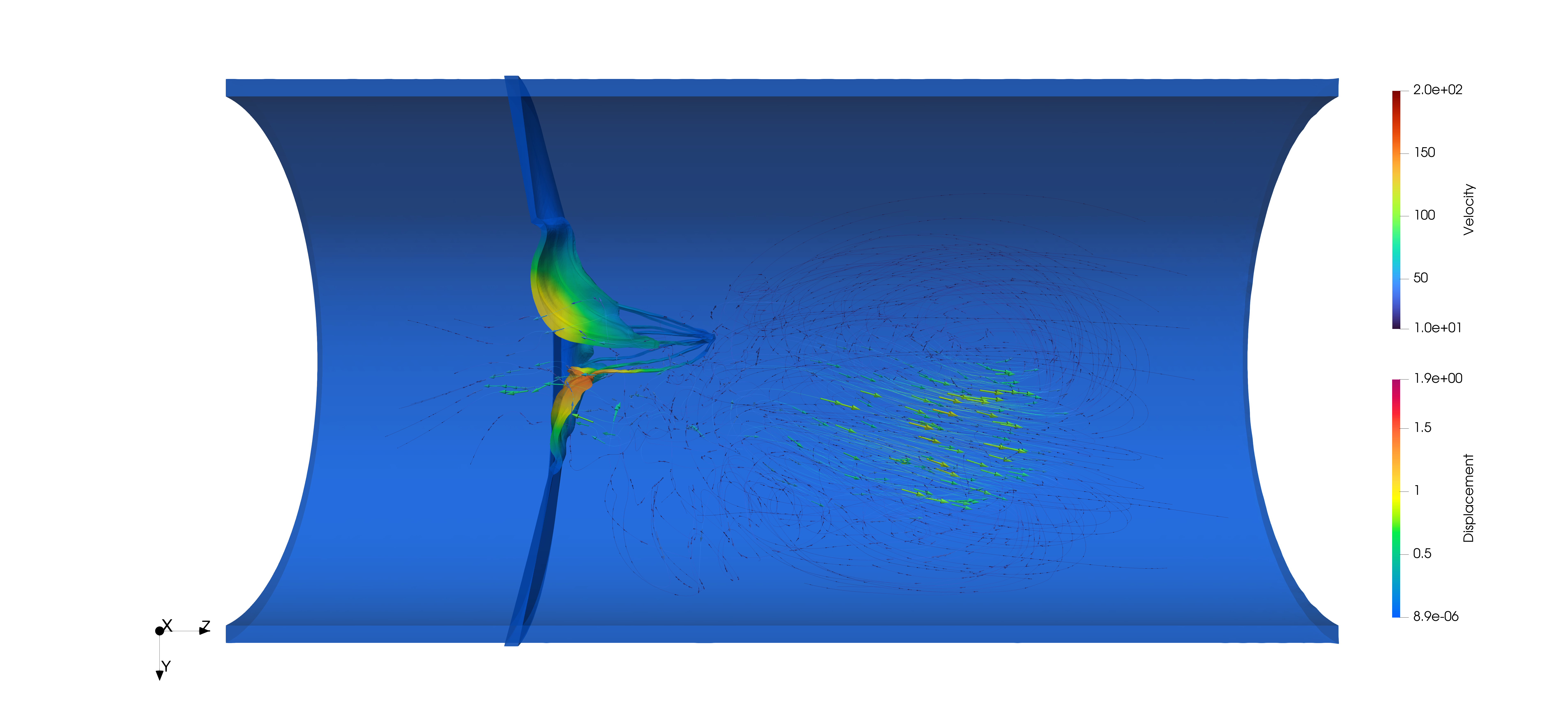}
            \subcaption{$0.22~\text{s}$}
            \label{fig::left_ventricle::b}
        \end{minipage}
        \vfill
        \begin{minipage}[t]{\textwidth}
            \centering
            \includegraphics[width=1.0\textwidth]{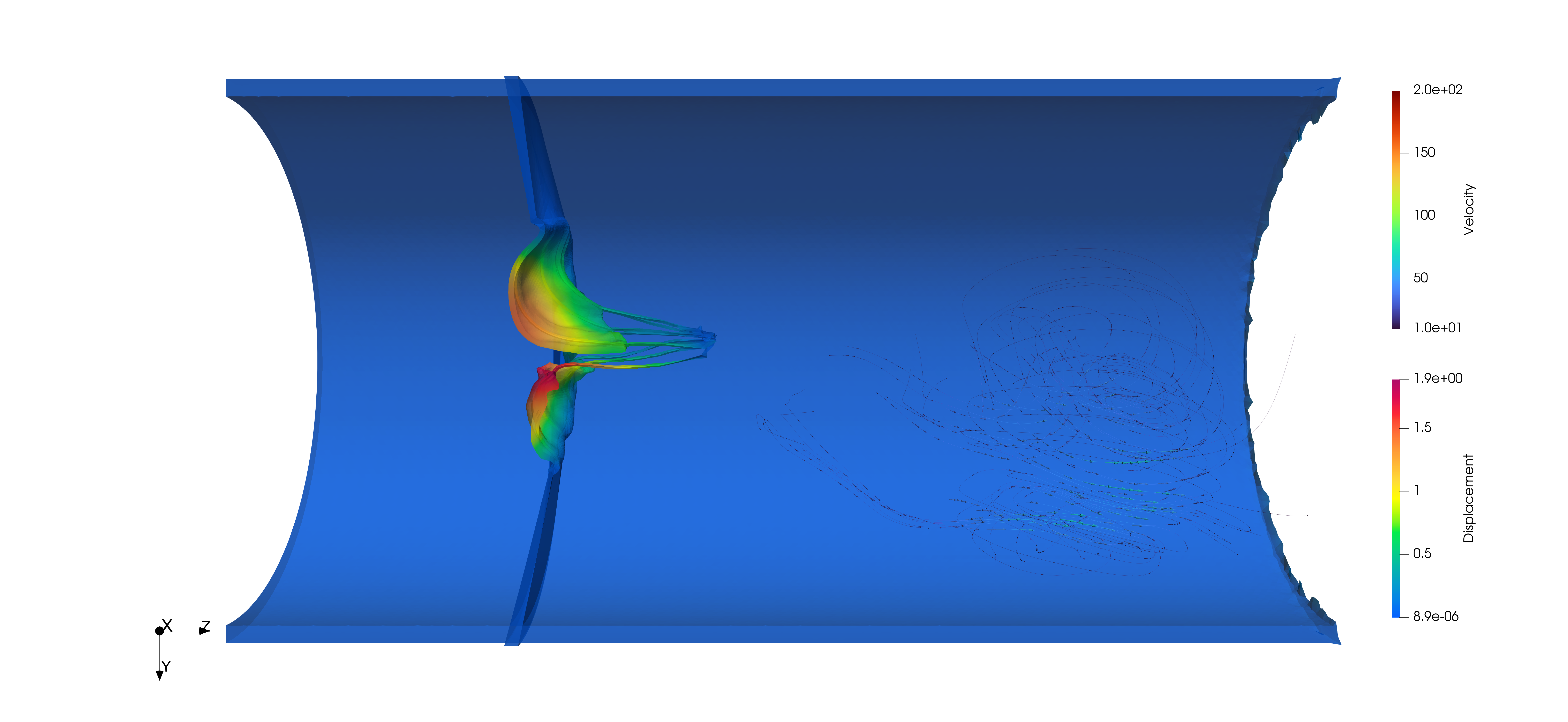}
            \subcaption{$0.72~\text{s}$}
            \label{fig::left_ventricle::c}
        \end{minipage}
    \end{minipage}
    \hfill
    \begin{minipage}[t]{0.48\textwidth}
        \centering
        \begin{minipage}[t]{\textwidth}
            \centering
            \includegraphics[width=0.8\textwidth]{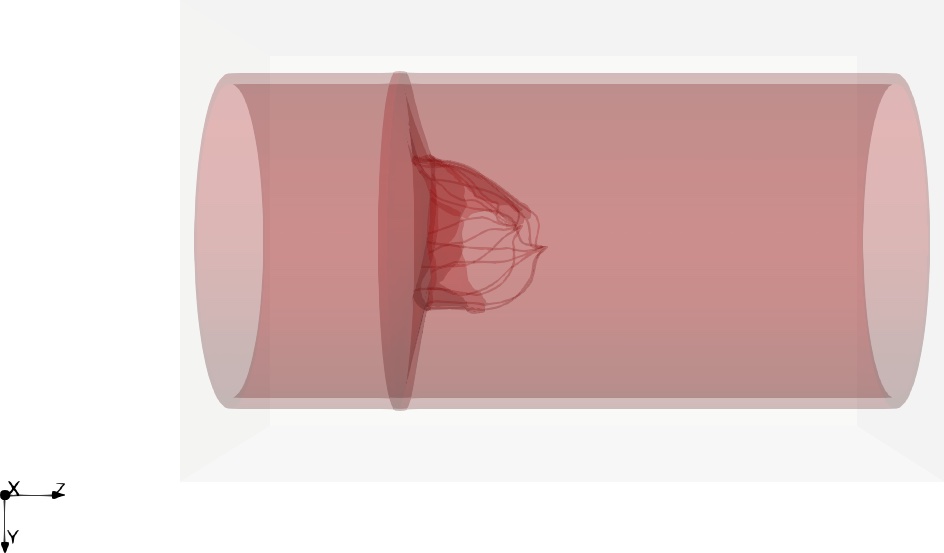}
            \subcaption{$0.10~\text{s}$}
            \label{fig::left_ventricle::d}
        \end{minipage}
        \vfill
        \begin{minipage}[t]{\textwidth}
            \centering
            \includegraphics[width=0.8\textwidth]{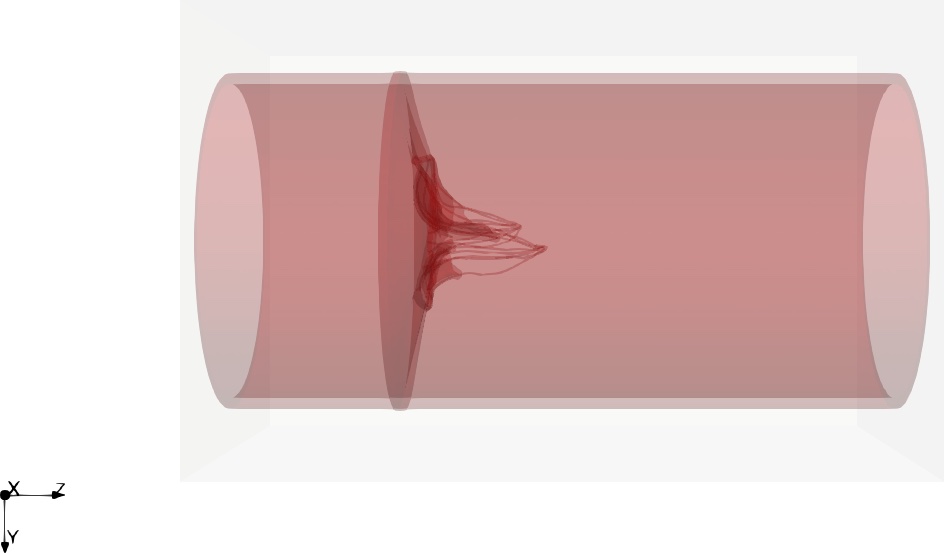}
            \subcaption{$0.22~\text{s}$}
            \label{fig::left_ventricle::e}
        \end{minipage}
        \vfill
        \begin{minipage}[t]{\textwidth}
            \centering
            \includegraphics[width=0.8\textwidth]{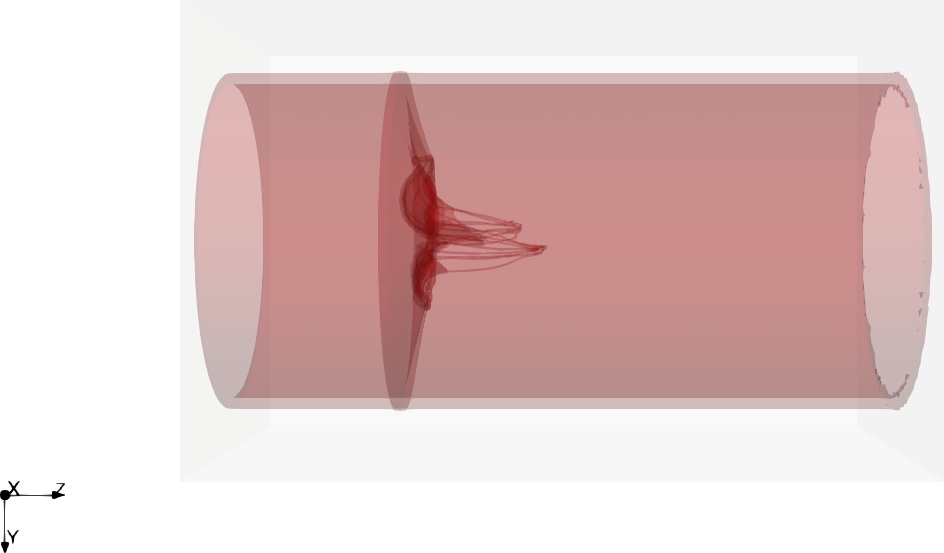}
            \subcaption{$0.35~\text{s}$}
            \label{fig::left_ventricle::f}
        \end{minipage}
    \end{minipage}
    \caption{Deformation of a left ventricular slice and fluid velocity streamlines at different times.}
    \label{fig::left_ventricle::combined}
\end{figure}

\begin{figure}[!ht]
    \centering
    \includegraphics[width=0.6\textwidth]{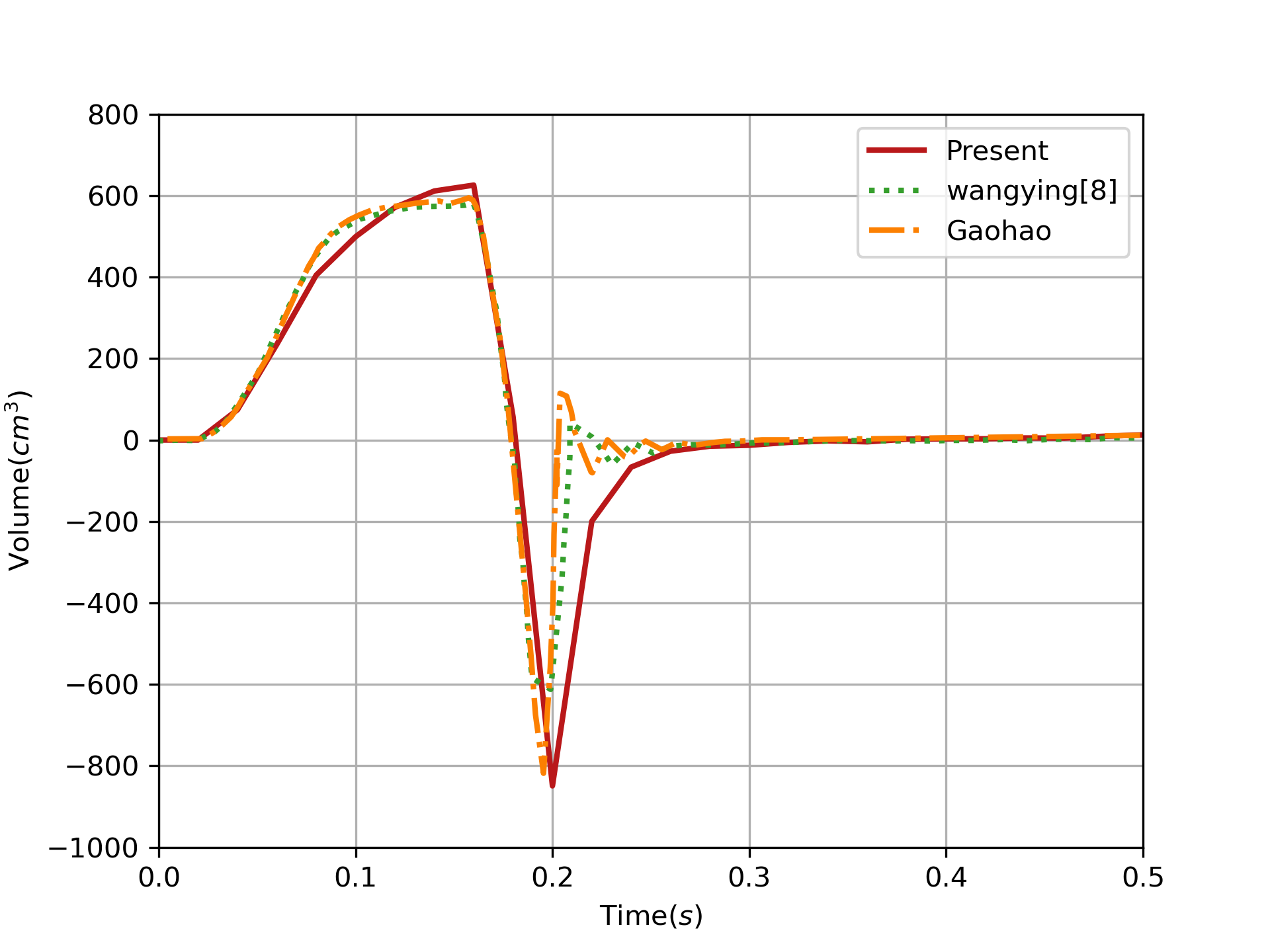}
    \caption{ }
    \label{fig::mitral_valve::model::flow_rate}
\end{figure}
This case does not involve the modeling of the left ventricle, so we only simulate the rapid filling phase during diastole and the systole phase. The pressure distribution for the simulated portion of the cardiac cycle is shown in Figure 4.


\section{Discussion and Conclusion}  
\label{sec:discussion}  

This study evaluates the performance of the nodal IB method on GPUs and explores its potential applications in cardiac mechanics.

\section*{Acknowledgments}
Xuan Wang, Li Cai and Pengfei Ma were supported by the National Natural Science Foundation of China Grants (\#12271440). 
Hao Gao was supported by the British Heart Foundation (PG/22/10930) and UK EPSRC (EP/S030875/1).

\bibliographystyle{unsrt}
\bibliography{references}  
\end{document}